\documentclass[prd,aps,a4,twocolumn,superscriptaddress,preprintnumbers,nofootinbib]{revtex4-1}

\usepackage{pslatex}
\usepackage[pdftex]{graphicx}
\usepackage{psfrag}
\usepackage{epsfig}
\usepackage{color}
\usepackage{cancel}
\usepackage{slashed}
\usepackage{amssymb}
\usepackage{amsmath}
\usepackage{hyperref}
\usepackage{enumerate}
\usepackage{multirow}
\usepackage{ulem}
\usepackage{float}
\usepackage{comment}
\usepackage{diagbox}

\allowdisplaybreaks

\begin{document}

\preprint{AIP/123-QED}

\title{Time variation of the atmospheric neutrino flux at dark matter detectors}

\author{Yi Zhuang}
\affiliation{Department of Physics and Astronomy, Mitchell Institute
  for Fundamental Physics and Astronomy, Texas A\&M University,
  College Station, TX 77843, USA}
\author{Louis E. Strigari}%
\affiliation{Department of Physics and Astronomy, Mitchell Institute
  for Fundamental Physics and Astronomy, Texas A\&M University,
  College Station, TX 77843, USA}
\author{Rafael~F.~Lang}  
\affiliation{Department of Physics and Astronomy, Purdue University, West Lafayette, IN 47907, USA}  
  
\begin{abstract}
The cosmic ray flux at the lowest energies, $\lesssim 10$ GeV, is modulated by the solar cycle, inducing a time variation that is expected to carry over into the atmospheric neutrino flux at these energies. Here we estimate this time variation of the atmospheric neutrino flux at five prospective underground locations for multi-tonne scale dark matter detectors (CJPL, Kamioka, LNGS, SNOlab and SURF). We find that between solar minimum and solar maximum, the normalization of the flux changes by $\sim 30\%$ at a high-latitude location such as SURF, while it changes by a smaller amount, $\lesssim 10\%$, at LNGS. A dark matter detector that runs for a period extending through solar cycles will be most effective at identifying this time variation. This opens the possibility to distinguish such neutrino-induced nuclear recoils from dark matter-induced nuclear recoils, thus allowing for the possibility of using timing information to break through the ``neutrino floor.''
\end{abstract}

\keywords{Cosmic Rays, Neutrino Floor, Atmospheric Neutrinos, Direct Dark Matter Detection}
\maketitle

\section{Introduction \label{sec:introduction}}

Dark matter detection experiments will soon be sensitive to neutrinos from the Sun, supernovae, and the atmosphere~\citep{Billard:2013qya,2019neuphys}. Turning their detection into information on properties of neutrinos and sources from which they originate requires to quantify the systematic uncertainties in the neutrino flux and interaction rate. The systematic uncertainty on the solar neutrino event rate has been a subject of numerous studies~\citep{Cerdeno:2017xxl}, including studies of its time-dependence~\cite{Davis:2014ama} and possible contributions from physics beyond the standard model~\citep{Dutta:2017nht,AristizabalSierra:2019ykk}. Understanding these systematics is especially important considering that xenon experiments are approaching sensitivity to the ${}^8$B component of the solar neutrino flux~\citep{Aprile:2020thb}. 

The systematic uncertainty on the atmospheric neutrino rate arises from several factors. One important factor is the normalization and the spectrum of cosmic rays impinging upon the Earth, which produce neutrinos via interactions in the atmosphere. The cosmic ray spectrum reaching the top of atmosphere is affected by solar modulation~\cite{Moraal:2013jxa,2013LRSP...10....3P,2013PAMELAproton,2018PAMELAproton}.  When local interstellar particles propagate through the heliosphere, they interact with the solar wind, experiencing diffusion and convection, accelerating or decelerating and drifting in the solar magnetic field. The structure of the heliosphere depends on the solar activity, which follows an 11-year cycle from one minimum to the next. A period of quiet solar activity allows for more incoming interstellar particles, and vice versa.

A second challenge to predicting the interaction rate of atmospheric neutrinos involves properly modeling the geomagnetic field. When low-energy \mbox{($\lesssim 10$ GeV)} cosmic rays approach the Earth, they are deflected by the geomagnetic field due to the rigidity cut-off. The rigidity cut-off determines which cosmic rays can enter the Earth, and it depends on direction for each location on Earth. This cut-off determines which cosmic rays are able to collide with nucleons in the atmosphere, generate extensive air showers, which produce mesons and leptons, which then interact and decay to produce atmospheric neutrinos. 

Including the effects of the geomagnetic field and the rigidity cut-off, three-dimensional calculations of the atmospheric neutrino flux have been performed~\citep{Honda:2001xy,Liu:2002sq,2005fluka,2003Wentz,Barr:2004br}. For cosmic ray energies \mbox{$\lesssim 10$ GeV}, three-dimensional calculations are especially important, and in particular for directions towards the horizontal, as from this direction the flux is enhanced relative to that from one-dimensional calculations~\citep{Lipari:2000wu}. The FLUKA calculations~\citep{20003DBattistoni} extend the atmospheric neutrino predictions to the lowest energies, \mbox{$\sim 10$ MeV}, where the dominant contribution is from pion and muon decay at rest. The FLUKA calculations are in good agreement with those from HKKM~\citep{Honda:2015fha} and Bartol~\citep{Barr:2004br} which are available for energies \mbox{$\gtrsim 100$ MeV}. 

The theoretical predictions for the low-energy atmospheric neutrino flux may be compared to  experimental measurements. The lowest energy measurements come from Super-Kamiokande (SK)~\citep{Richard:2015aua} and from Frejus~\citep{Daum:1994bf}. SK detects neutrinos via electrons and muons that are produced in and around the detector. SK is sensitive to Charged Current interactions, and is able to distinguish between flavors due to the nature of the events produced in the detector. These results provide a measurement of the neutrino flux down to energies $\gtrsim 100$ MeV. At the energies studied by SK, the flux normalization is consistent with theoretical predictions at these energies~\citep{Honda:2015fha}. The Sudbury Neutrino Observatory (SNO) and KamLAND are also sensitive to atmospheric neutrinos in this energy range~\citep{Aharmim:2020agi,Abe:2021tkw}. 

In spite of the experimental and theoretical progress in measuring the atmospheric neutrino flux for \mbox{$\gtrsim 100$ MeV} energies, there are no direct measurements at lower energies. However, it is these neutrino energies that future direct dark matter detection searches are sensitive to~\citep{Newstead:2020fie}. Measuring the flux in this energy regime is thus especially important. In addition, measurement of atmospheric neutrinos at these energies has implications for the  DUNE experiment~\citep{Kelly:2019itm,Kelly:2021jfs,Denton:2021rgt} and JUNO~\citep{Abusleme:2021paq}. 

\begin{figure*}[!htbp]
    \includegraphics[width = 0.45\textwidth]{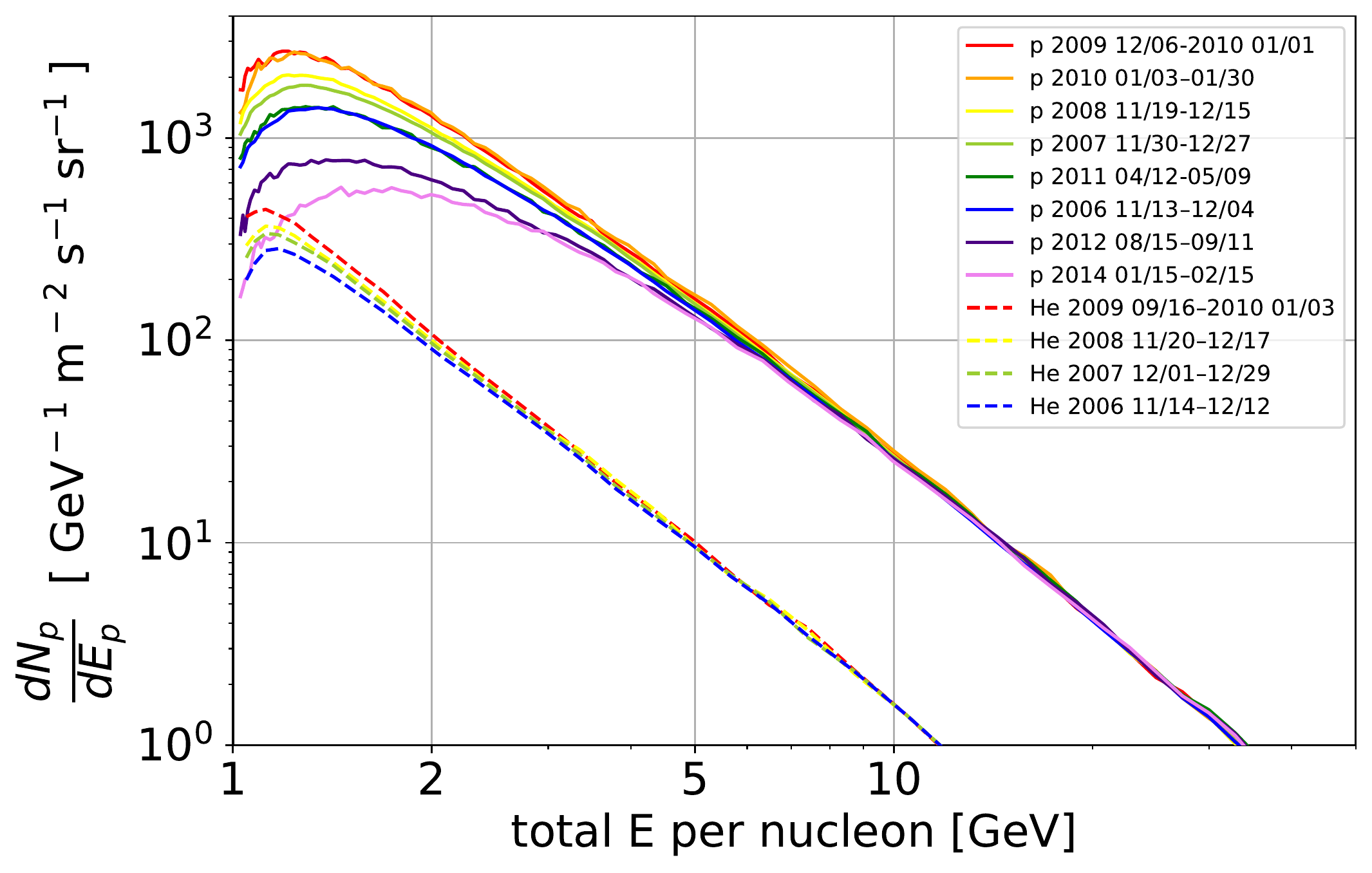}
    \includegraphics[width = 0.45\textwidth]{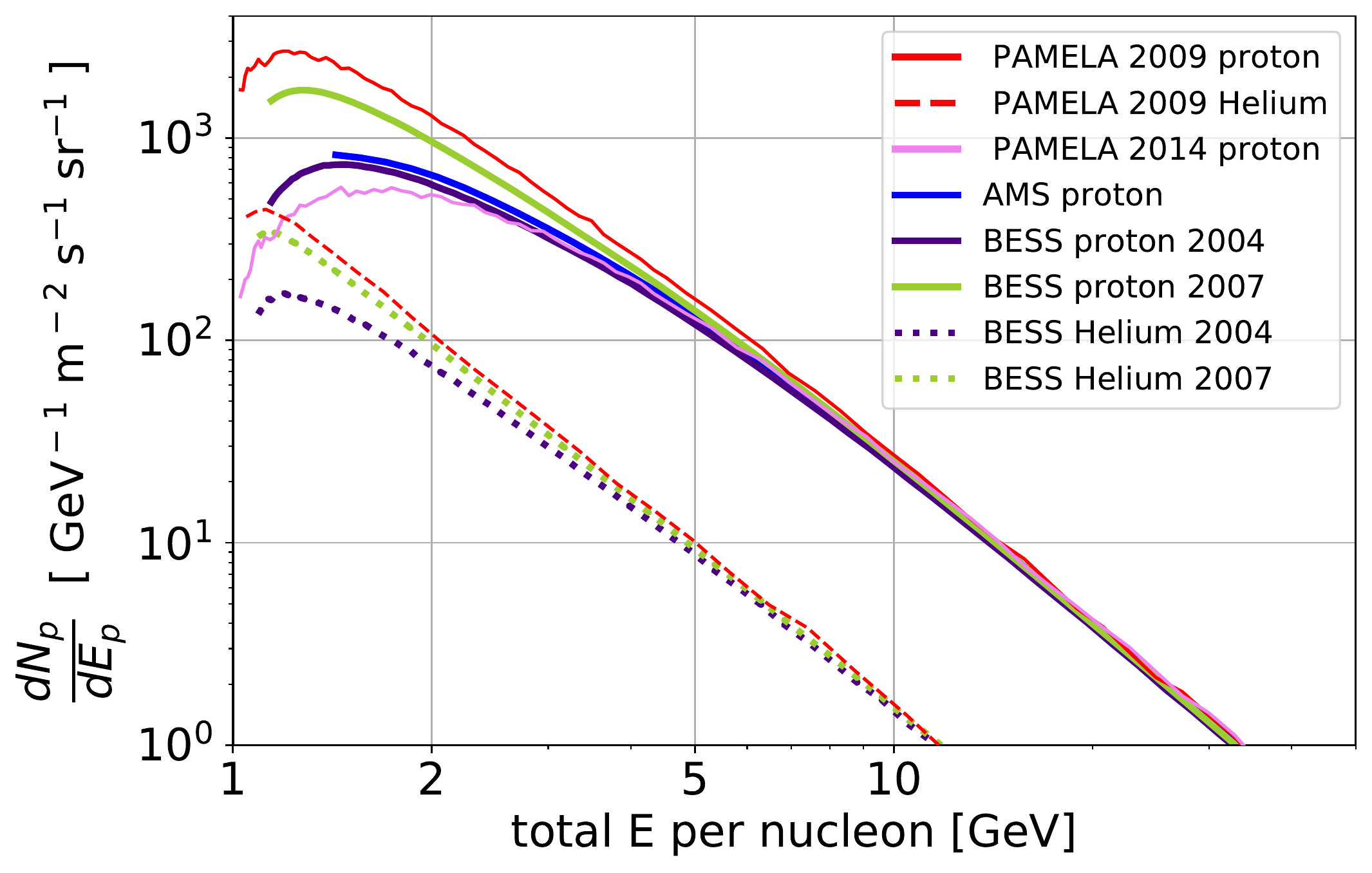}
    \caption{Left: Primary cosmic ray proton (solid curves) and helium (dashed curves) spectra plotted as a function of total energy per nucleon, as measured by PAMELA for the years indicated~\cite{2018PAMELAproton}. Right: Primary cosmic ray proton and helium spectra as measured by BESS~\citep{Abe:2015mga} for the years indicated, and the spectra measured by AMS~\citep{2015PhRvL.114q1103A}, in comparison to the minimum and maximum fluxes from PAMELA.
    \label{fig:protonflux}}
\end{figure*}

In this paper, we study the atmospheric neutrino flux at low energies, and use these results to better understand systematics in direct dark matter detection searches. We particularly focus on the time-dependence of the atmospheric neutrino flux due to solar modulation, and quantify this time variation at five possible detector locations, including the Laboratori Nazionali del Gran Sasso (LNGS) and the Sanford Underground Research Facility (SURF). We explore the prospects for measuring this time variation at each detector location. 

This paper is organized as follows. In Section~\ref{sec:CRflux}~we discuss the properties of the cosmic ray flux at the energies that we are interested in. In Section~\ref{sec: protontonu} we detail how the neutrino flux is extracted from the primary cosmic ray flux at each location. Section~\ref{sec: scattering}~reviews the calculation of the event rate at a dark matter detector through the coherent elastic  neutrino-nucleus scatter (CE$\nu$NS) process. In section~\ref{sec:results} we present our predictions for the event rate before discussing and concluding in section~\ref{sec:conclusions}. 

\section{Cosmic Ray Flux} 
\label{sec:CRflux}
In this section, we review the measurements and discuss the properties of the cosmic ray (CR) flux that will be most relevant for our analysis. We discuss our estimation of the CR flux at each detector location, focusing on how we handle the effect of solar modulation and the geomagnetic rigidity cut-off. 

CRs are produced by the acceleration of charged particles in Galactic and extragalactic environments~\citep{Gaisser:2016uoy}. CRs are observed over a vast range of energies, from \mbox{$\sim 100$ MeV} to upwards of \mbox{$10^{20}$ eV}. The dominant component of CRs are protons. After protons, the next most significant component is helium, with a flux of $\sim 10\%$ that of protons~\citep{2015PhRvL.115u1101A,Abe:2015mga,Marcelli:2020uqv}. Heavier elements make progressively smaller contributions to the flux~\citep{2017PhRvL.119y1101A}. Due to their relatively small fluxes we do not consider elements heavier than helium in our analysis.  

Above energies of \mbox{$\gtrsim 10$ GeV}, the primary proton flux measured on Earth is described by a power law, \mbox{$dN/dE = {\rm Norm} (E/{\rm GeV})^\gamma$}, where $\gamma \approx -2.74$~\citep{AMS:2021nhj} and Norm is the flux normalization at 1 GeV. Below these energies, due to diffusion through the solar wind~\citep{Gleeson:1968zza} and the geomagnetic field, the proton spectrum measured on Earth differs from that in the local interstellar medium~\citep{Cummings:2016pdr}. Due to these effects, for energies \mbox{$\lesssim 10$ GeV}, the measured proton spectrum flattens. CR energies \mbox{$\gtrsim 10$ GeV} are not affected by solar modulation. The primary helium flux has been measured at solar minimum for energies per nucleon up to \mbox{20 GeV}~\citep{Abe:2015mga,Marcelli:2020uqv}, and is observed to have the same spectral shape as protons. 

The CR spectra as measured by the PAMELA~\citep{2013PAMELAproton,2018PAMELAproton}, BESS~\citep{Abe:2015mga}, and AMS~\citep{2015PhRvL.114q1103A} are shown in Figure~\ref{fig:protonflux}. The PAMELA measurements are presented at the most number of epochs, showing the variation over the entire range from solar minimum to solar maximum. For both protons and helium, the minimum and maximum spectra are consistent with the BESS and the AMS spectra. Power law fits to the proton spectra are given below. For helium, we fit power laws to two total energy bins: \mbox{$[4-4.8]$ GeV}, and \mbox{$[4.8-400]$ GeV}, where \mbox{4.8 GeV} is the approximate turnover point of the observed helium spectrum. The power law slopes in these two energy ranges are $[0.15, -2.65]$, respectively, and the normalizations are $[2.8, 2.8]$ $\mathrm{GeV^{-1}cm^{-2}s^{-1}sr^{-1}}$ per particle.

We consider several different prospective detector locations at which we determine the CR flux and the atmospheric neutrino flux. These locations are listed in Table~\ref{tab:mag}, with their corresponding geographic and geomagnetic coordinates. All of these locations represent possible locations for next generation dark matter detectors. The calculations for Kamioka compare to previous estimates of the atmospheric neutrino flux. Also shown in Table~\ref{tab:mag} are the horizontal ($B_x$) and vertical ($B_z$) components of the magnetic fields at these locations. The horizontal component is defined to point towards the magnetic north, and the vertical component points downward towards the center of the Earth. With this notation, the magnetic field component in the orthogonal direction pointing towards the magnetic west is \mbox{$B_y = 0$}. Values of the $B-$field are computed using the IGRF13~\cite{2020IGRF13} from~\cite{GEOMAG} at an elevation of $56.4$ km above mean sea level (half of distance from top of atmosphere 112.8 km defined in CORSIKA) for the dates shown in Table~\ref{tab:mag}. These dates approximately correspond to the measurements of the CR flux that are shown in Figure~\ref{fig:protonflux}.

\begin{table*}[!htbp]
    \centering
    
    \begin{tabular}{cccccc}
    \hline
    Location & Geographic Coordinate & Date & Geomagnetic Coordinate & $B_x$ ($\mu$T) & $B_z$ ($\mu$T)\\
    \hline
    \multirow{2}{*}{CJPL} &\multirow{2}{*}{ 28.15323$^{\circ}$N, 101.7114$^{\circ}$E}  & 2009/12/18 & 18.06$^{\circ}$N, 174.36$^{\circ}$E &  34.7213  & 32.4124 \\
     &&2014/01/30 & 18.29$^{\circ}$N, 174.65$^{\circ}$E & 34.5752  &  32.8396\\ 
   
    \multirow{2}{*}{Kamioka} &\multirow{2}{*}{36.4267$^{\circ}$N, 137.3104$^{\circ}$E}  & 2009/12/18 & 27.43$^{\circ}$N, 153.40$^{\circ}$W & 29.2654 & 35.6555\\
     &&2014/01/30 & 27.66$^{\circ}$N, 153.06$^{\circ}$W & 29.2440 & 35.8401\\ 
   
    \multirow{2}{*}{LNGS}&\multirow{2}{*}{42.4531$^{\circ}$N, 13.5739$^{\circ}$E } & 2009/12/18 & 42.20$^{\circ}$N, 94.88$^{\circ}$E & 23.5180 & 38.6319\\
    && 2014/01/30 & 42.17$^{\circ}$N, 94.98$^{\circ}$E & 23.5695 & 38.7400 \\
    
    \multirow{2}{*}{SURF}&\multirow{2}{*}{44.3517$^{\circ}$N,   103.7513$^{\circ}$W}& 2009/12/18 & 52.38$^{\circ}$N, 37.94$^{\circ}$W &  18.0779 &  50.6775 \\
    &&2014/01/30 & 52.23$^{\circ}$N, 37.38$^{\circ}$W & 18.0778 & 50.1562\\
    
    \multirow{2}{*}{SNOlab}&\multirow{2}{*}{46.4733$^{\circ}$N,   81.1854$^{\circ}$W}& 2009/12/18 & 56.12$^{\circ}$N, 11.15$^{\circ}$W & 	16.3162 & 51.9549\\
    &&2014/01/30 & 55.89$^{\circ}$N, 10.69$^{\circ}$W & 16.4926  & 51.4179 \\
    \hline
    \end{tabular}
    \caption{Properties of the magnetic fields at the detector locations with geomagnetic coordinates that we consider. The field is quoted 56.4 km above mean sea level. The values listed are from the IGRF13 model, with $B_x$ (the horizontal component) pointing to magnetic north, and $B_z$ (the vertical component) pointing downward towards the center of the Earth.~\label{tab:mag}}
\end{table*}

To provide an understanding of the impact of the geomagnetic field on the CR flux and then ultimately on the neutrino flux, we consider two models, labeled here Stoermer and track-back, respectively. The first is a simple model which assumes that the CR spectrum is cut-off below a specific rigidity, defined as \mbox{$R = pc/Ze$}, where $p$ is the momentum, $Z$ is the proton number, and $e$ is the charge. The rigidity cut-off is given by the Stoermer formula~\citep{1991NCimC..14..213C}, which depends on the incoming zenith ($\theta$) and azimuthal ($\phi$) angles of the CR and the geomagnetic latitude ($\lambda$) as
\begin{equation}
    \label{eqn: stoermer}
    R = 59.4 \, {\textrm{GeV}} \frac{\cos^{4}\lambda}{r^{2}\left[1+(1-\cos^{3}\lambda \sin\theta \sin\phi)^{1/2}\right]^{2}}. 
\end{equation}
Here the azimuthal angle is defined clockwise from the $x$-axis and increases towards the east, and $r$ is the distance from the center of the Earth. 

Though the Stoermer formula is important in determining the effect of a dipolar magnetic field on the CR spectrum at a specific location, and for providing a physical intuition into the impact of the rigidity cut-off, there are limitations due to simplifying assumptions. The first assumption is that of a dipolar model itself, which is different than the true geomagnetic field. Second, it does not account for CR diffusion in the penumbra region. This is the region which, at a fixed rigidity, distinguishes between the forbidden and allowed regions for CRs of all incoming trajectories~\citep{1991NCimC..14..213C}. CRs diffuse in this penumbra region resulting in a more complex fine-grained structure in the CR spectrum. Third, it does not account for the shadow of the Earth, in which CR trajectories are on Earth-crossing orbits~\cite{stanev2010high}.

\begin{figure}[!htbp]
    \centering
    \includegraphics[width =0.5 \textwidth]{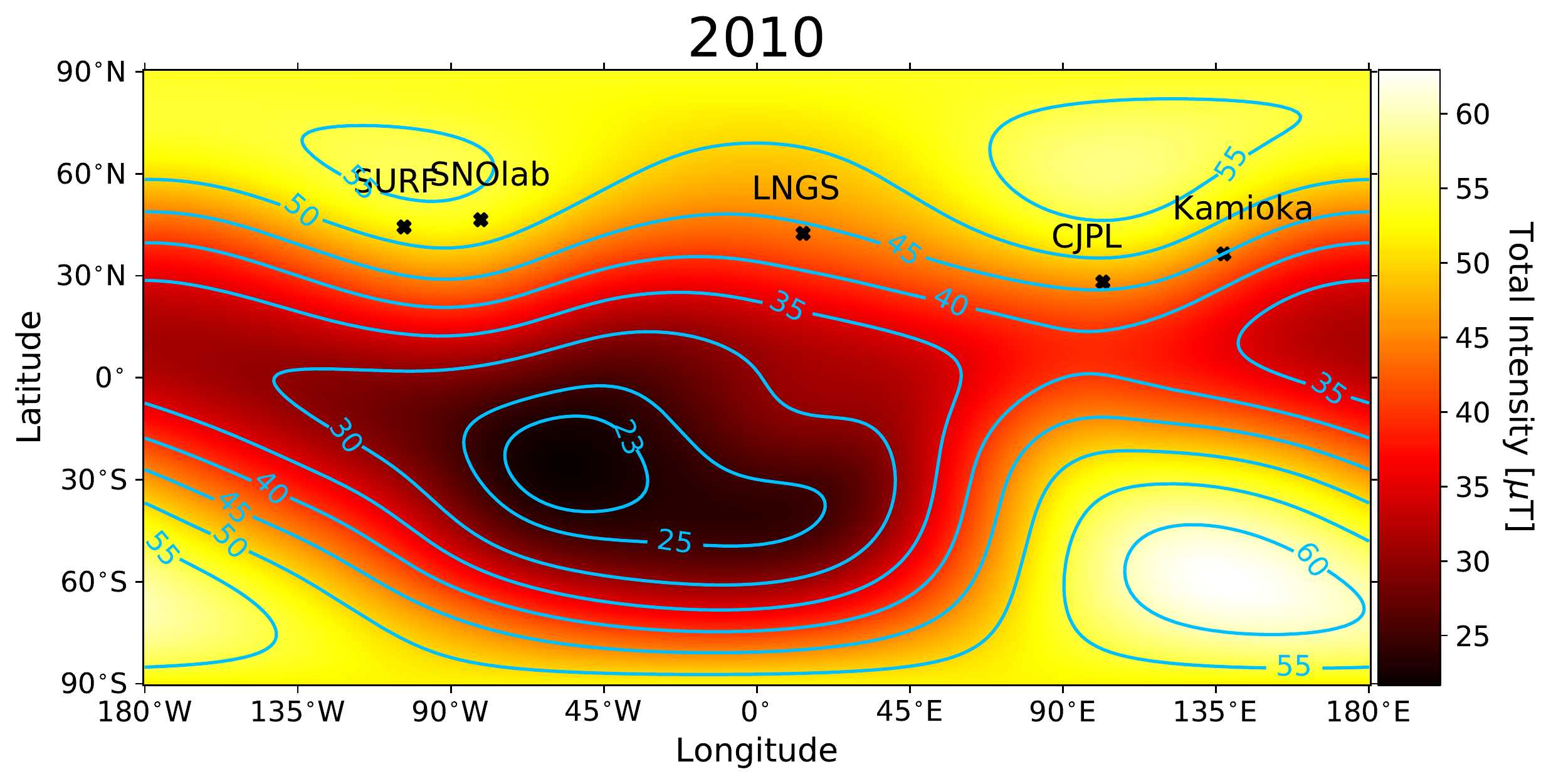}
    \caption{The geomagnetic model IGRF/DGRF13. The field is shown for solar minimum during 2010. The locations of the underground laboratories considered here are indicated.}
    \label{fig:Bfield}
\end{figure}

To compare to the Stoermer model for the rigidity cut-off, we consider the evolution of charged particles in a more realistic magnetic field. This ``track-back" method uses the IGRF13 magnetic field model and boosts protons and helium nuclei with negative charge outward from the top of the atmosphere, which we define as \mbox{$h = 112.8$} km~\cite{Heck:1998CORSIJKA}, using the equation of motion, \mbox{$d(\gamma m {\vec v})/dt = q {\vec v} \times {\vec B}$}. The IGRF13 magnetic field is obtained using the code from IAGA SummerSchool2019~\cite{IAGA_code}, with the strength of the field shown in Figure~\ref{fig:Bfield} in 2010 for solar minimum.

To solve the equation of motion, we numerically integrate to solve for the position and velocity of the particle, with the $\vec B$-field updated at each step as the particle evolves. Once launched, the particle either oscillates, falls back to the Earth, or escapes. It is also possible that the particle touches the surface of the Earth, in which case the algorithm terminates. If the calculation reaches a timescale equivalent of 15 seconds with its position reaching more than 30 $R_{\oplus}$ and total distance less than 500 $R_{\oplus}$, the particle is considered as escaped. The cutoff values 30 $R_{\oplus}$ and 500 $R_{\oplus}$ are adopted from previous studies~\cite{1995Lipari,stanev2010high}. Practically, particles hit the Earth or escape rapidly, and we find 15 seconds is a sufficient  computational time to check whether a particle continues oscillating or escapes after several oscillation cycles.   

To obtain initial positions and momenta for the protons and helium nuclei used in the track-back model, we used the values for the CRs sampled from the CORSIKA simulation, which we describe in more detail in the following section. For the purposes of the discussion in this section, CORSIKA is used to simulate CRs uniformly over zenith angles of $[0,90]$ degrees and azimuthal angles within $[0, 360]$ degrees. From these initial conditions, we determine whether the proton or helium nucleus can escape or is trapped given the IGRF13 geomagnetic model. 

We bin the particles according to their energies, with the exact energy binning depending on the detector location. The corresponding total energy ranges (in GeV) for each detector location are as follows: $[5,10]$, $[10,15]$, $[15,20]$,  $[20,40]$, $[40,60]$ for CJPL; $[5,10]$, $[10,15]$, $[15,20]$, $[20,40]$ for Kamioka; $[2,5]$, $[5,10]$, $[10,15]$, $[15,20]$ for LNGS; $[1.3,2]$, $[2,5]$, $[5,10]$ for SURF and $[1.3,2]$, $[2,5]$ for SNO. In each of the energy ranges above, we generate 1500 protons for backtracking. This then implies that a total of 6000 protons are used for testing at Kamioka and LNGS, 7500 for CJPL, 4500 for SURF, and 3000 for SNO. For helium the total energy ranges are $[4,4.8]$, $[4.8,80]$, $[80,160]$, $[160,240]$, $[240,320]$, $[320,400]$, which is wider than for protons, because the measured flux for helium is given in units per nucleon and as above \mbox{4.8 GeV} is the approximate helium particle total energy where the flux starts to turn over. Tracking back is applied to energy ranges $[4.8,80]$ and $[80,160]$ for 3000 helium nuclei at CJPL, and $[4.8,80]$ for 1500 helium nuclei at the remaining locations. For each respective detector location, all CRs are rejected below the minimum energy indicated and all CRs escape above the maximum indicated energy. 

\begin{figure*}[!htbp]
    \centering
    \includegraphics[width = \textwidth]{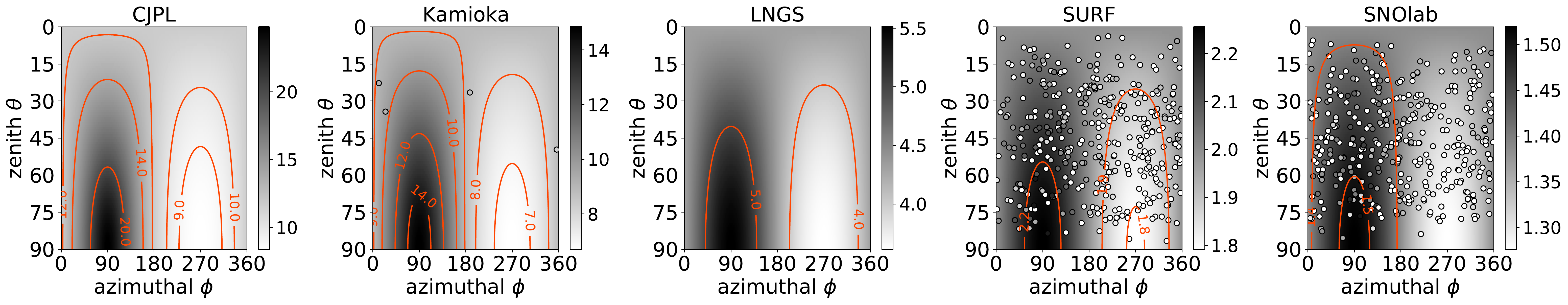}
    \includegraphics[width = \textwidth]{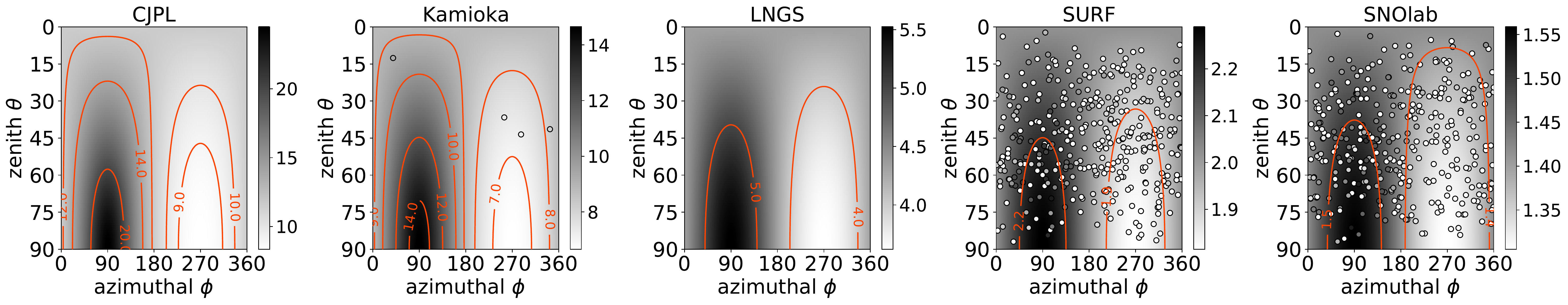}
    \caption{Rigidity cut-offs in the Stoermer and trackback models for each detector location. Top row is for 2009, and bottom row is for 2014. All detectors are in geomagnetic coordinates, with the grey-scale background representing the Stoermer rigidity within zenith angles $[0,90]$ degrees and azimuthal angles $[0, 360]$ degrees, with the azimuthal angle measured clockwise from magnetic north. Several example contours indicated in red. Dots are protons less than Stoermer rigidity but escape in the trackback method. 
    \label{fig:stomer_trackback}}
\end{figure*}

In Figure~\ref{fig:stomer_trackback}, we show the results for the rigidity cut-off calculation for both the Stoermer and trackback models, for several primary proton energy ranges. Shown as dots are the particles that have momentum less than the Stoermer rigidity but can escape in the trackback method. For low-latitude locations, a small number of particles escape in the trackback method below the Stoermer rigidity, implying that both models give a consistent estimate of the rigidity cut-off. An exception is at Kamioka, where the large penumbra width at rigidity \mbox{$\approx$ 9 GeV} allows several particles to escape. On the other hand, for high-latitude locations, many particles escape in the trackback method below the Stoermer rigidity, implying that the low-energy cosmic ray flux at these locations is more model-dependent. We note that we use the 2015 geomagnetic field for solar maximum, which is indistinguishable from the 2010 model that we use for solar minimum. These are the closest available geomagnetic models to the years of data we study, which are 2009 for solar minimum to include helium and 2014 for solar maximum.

\begin{figure*}[!htbp]
    \includegraphics[width = \textwidth]{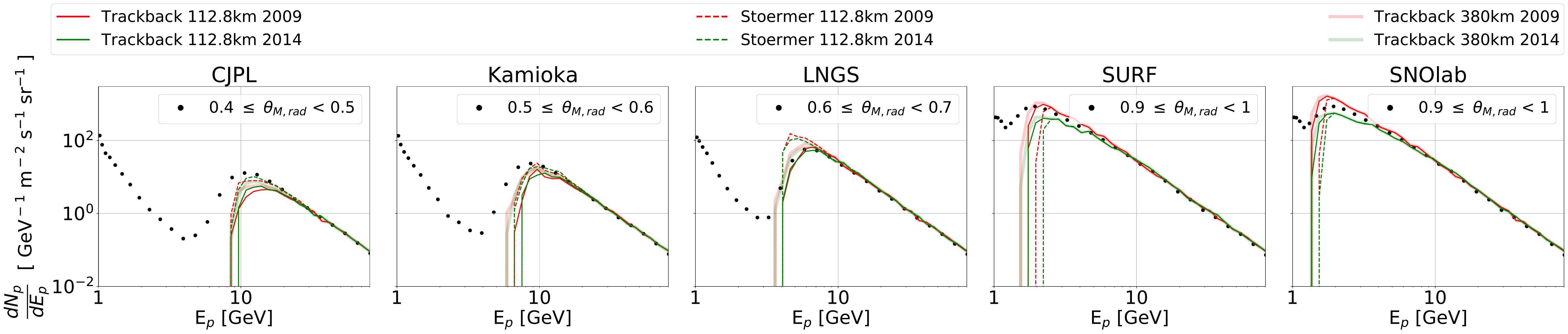}
  
    \caption{Cosmic ray proton flux after applying both the Stoermer formula and the trackback geomagnetic model at 112.8 km and 380 km altitudes, for each detector location. The geomagnetic latitude $\mathrm{\theta_{M, rad}}$ at each location is obtained from AACGM-v2~\cite{aacgm_online} for each date in Table~\ref{tab:mag}. For comparison, the observed proton flux as measured by AMS at 380 km altitude is shown at different geomagnetic latitudes~\cite{2000AMS} (black dots). }
    \label{fig:proton_flux_after_rigidity}
\end{figure*}

The resulting CR fluxes at each location are shown in Figure~\ref{fig:proton_flux_after_rigidity}. The power law fits and the normalizations over the various energy ranges are shown in Table~\ref{tab: slope_norm}. The geomagnetic latitude ($\theta_\mathrm{M}$) for comparison at each location is at 112.8 km altitude using Altitude Adjusted Corrected Geomagnetic Coordinates (AACGM) geocentric, which is different from those in Table~\ref{tab:mag}, and also varies among versions of AACGM and years. We use AACGM-v2~\cite{2014AACGM2} to match time variations, and \mbox{$\theta_\mathrm{M}$ $\approx$ 22$^{\circ}$} at CJPL, 30$^{\circ}$ at Kamioka, 37$^{\circ}$ at LNGS, 54$^{\circ}$ at SURF and 56$^{\circ}$ at SNO. The shapes of the spectra that we calculate are in agreement with the proton flux measurements from AMS at \mbox{380 km} over the respective ranges of magnetic latitude~\citep{2000AMS}. The increased flux at low energy comes from protons being generated in the atmosphere and inner radiation belt for the region near the South Atlantic Anomaly. The peak difference is due to different versions of geomagnetic coordinate systems. Small shape differences at the turnover energy are due to the diffusion pattern from the penumbra region, which requires high angular resolution to accurately map out.  For example, with \mbox{$250 \, \mu$sr} resolution~\cite{2003Wentz}, the diffusion in simulation agrees with the AMS data~\cite{2000AMS}. 

\begin{figure*}[!htbp]
    \includegraphics[width = \textwidth]{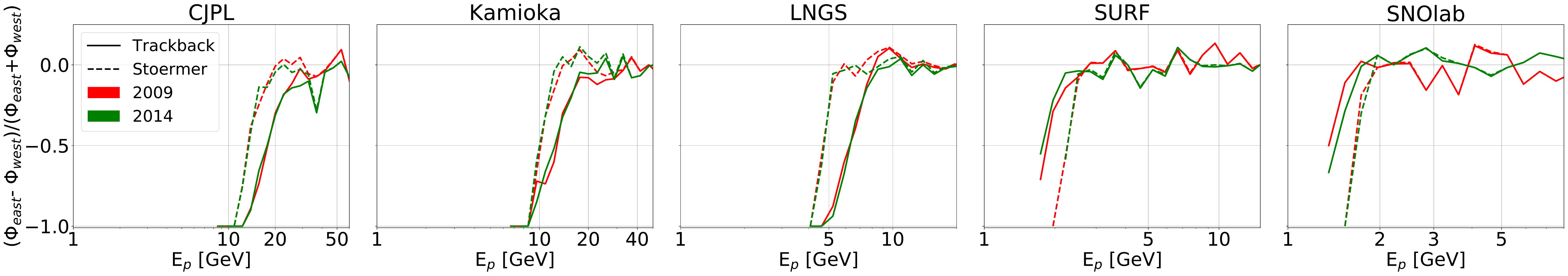}
    \caption{East-west asymmetry in the flux as a function of proton energy for solar minimum and solar maximum, for the Stoermer and trackback geomagnetic models.}
    \label{fig:eastwest}
\end{figure*}

The rigidity is higher in the east, meaning that fewer particles travel from east, so more particles penetrate from the west and are recorded towards the east~\citep{Lipari:2000du}. Such an east-west asymmetry is visible in Figure~\ref{fig:eastwest}. Shown is the ratio of the difference in the flux from the east, $\phi_\mathrm{east}$, to that of from the west, $\phi_\mathrm{west}$, divided by the sum of the fluxes. The symmetry is more evident at larger energies for low latitude locations, and at smaller energies for high latitude locations. Though the asymmetry is clearly visible, due to the energy and angular resolution of our simulations we do not distinguish between the trackback and Stoermer models using the east-west flux. Already here one can see that the higher-latitude locations (such as SURF and SNOlab) receive more low-energy flux, which exhibits stronger modulation. This effect carries through our study. 

\begin{table}[!htbp]
    \caption{Proton slopes and normalizations in the energy bins used at the detector locations. The ``$/$" indicates that no neutrinos are produced within this energy range because protons are deflected by the geomagnetic field. For \mbox{$\mathrm{E_p}$ $\gtrsim 10$ GeV}, the slope is \mbox{$\gamma \approx -2.74 $}. The units of Norm are $\mathrm{GeV^{-1} cm^{-2}s^{-1}sr^{-1}}$.}
    \centering
    \begin{tabular}{cccccc}
    \hline\hline
    Proton & \backslashbox{Year}{$\mathrm{E_p}$ [GeV]} & 1-1.3 & 1.3-2 & 2-5 & 5-10\\
    \hline\hline
     \multirow{2}{*}{Slope ($\gamma$) } & 2009 & 1.68 & -1.7 & -2.27 & -2.58\\
    & 2014 & 4.34 & 0.17 & -1.46 & -2.27\\
    \hline
    \multirow{10}{*}{Norm} &2009 Kamioka &/&/&/& 1.1\\
    &2009 CJPL &/&/&/& 1.1\\
    &2009 LNGS  &/&/&  0.65 & 1.1\\
    &2009 SURF &/& 0.4&  0.65 & 1.1\\
    &2009 SNOlab &/& 0.4&  0.65 & 1.1\\
    & 2014 CJPL &/&/&/& 0.5\\
    & 2014 Kamioka &/&/&/& 0.5\\
    &2014 LNGS  &/&/& 0.15 & 0.5\\
    &2014 SURF &/& 0.05 & 0.15 & 0.5\\
    &2014 SNOlab  &/& 0.05 & 0.15 & 0.5\\
    \hline
    \end{tabular}
    \label{tab: slope_norm}
\end{table}

\section{Atmospheric Neutrino Flux \label{sec: protontonu}}

We now move on to estimating the atmospheric neutrino flux and its associated time variation from CR interactions. We must estimate the flux at several detector locations that do not have flux measurements at low energies in the published literature. We begin by briefly describing the CORSIKA code for simulating the atmospheric neutrino flux. We then describe the modifications and additions to CORSIKA that are required for our analysis, and then present the estimates for the neutrino flux and the time variation at each detector location. 

The CORSIKA program~\cite{1998corsika} generates neutrinos from simulations of CR interactions and the subsequent air showers. Within CORSIKA, we use the FLUKA model to simulate low energy events, \mbox{$< 80$ GeV}, and QGSJET 01C for higher energy events. For the detector, we use a horizontal flat detector array at sea level. We simulate primary particles over a zenith range from $[0,90]$ degrees, and track the decay modes of particles using the EHISTORY option. The input geomagnetic latitude for each detector is in Table~\ref{tab:mag} and \mbox{$r=\frac{h+R_{\oplus}}{R_{\oplus}}$}, where \mbox{$h=112.8 $ km}. For our generated neutrino events, we do not include neutrino oscillations, since our detection is via a neutral current process that is flavor-independent. 

We calculate the neutrino spectrum from primary particles in the energy ranges indicated above as (e.g.~\cite{2016Schoneberg}):
\begin{equation}
    \label{eqn: CRtoneutrino}
    \phi(E_{\nu}) = \frac{N_{\nu}}{\Delta E_{\nu}} \frac{s_\mathrm{A}}{N_\mathrm{shower}} \int_{E_\mathrm{CR, min}}^{E_\mathrm{CR, max}} \Phi_\mathrm{CR} dE_\mathrm{CR}. 
\end{equation}
Here $N_\nu$ is number of neutrinos within the input energy bin, \mbox{$N_\mathrm{shower} = 1500$} for each energy range, and \mbox{$s_\mathrm{A}$ = 1.018} accounts for the area difference between the top of atmosphere and the detector~\citep{2003Wentz}. The input CR flux, $\Phi_\mathrm{CR}$, represents either the primary proton or helium flux, weighted by their contributions to the total flux, and the energies $E_\mathrm{CR, min}$ and $E_\mathrm{CR, max}$ correspond to the minimum and maximum energies for the input primaries. The primary particle energy runs up to \mbox{400 GeV}. Since the rigidity cut-off varies among the different detector locations, each location starts with a different $E_\mathrm{CR, min}$ as is listed in Table~\ref{tab: slope_norm}. 
We use linear neutrino energy bins in log space, with $50$ bins from \mbox{$10$ MeV} to \mbox{$3$ GeV}; this bin size is chosen to be similar to that used in the FLUKA simulations~\citep{20003DBattistoni} over this same energy range, with $\Delta \log_{10}E_\nu \approx 0.05$. 

\begin{figure}[!htbp]
    \centering
    \includegraphics[width = 0.45\textwidth]{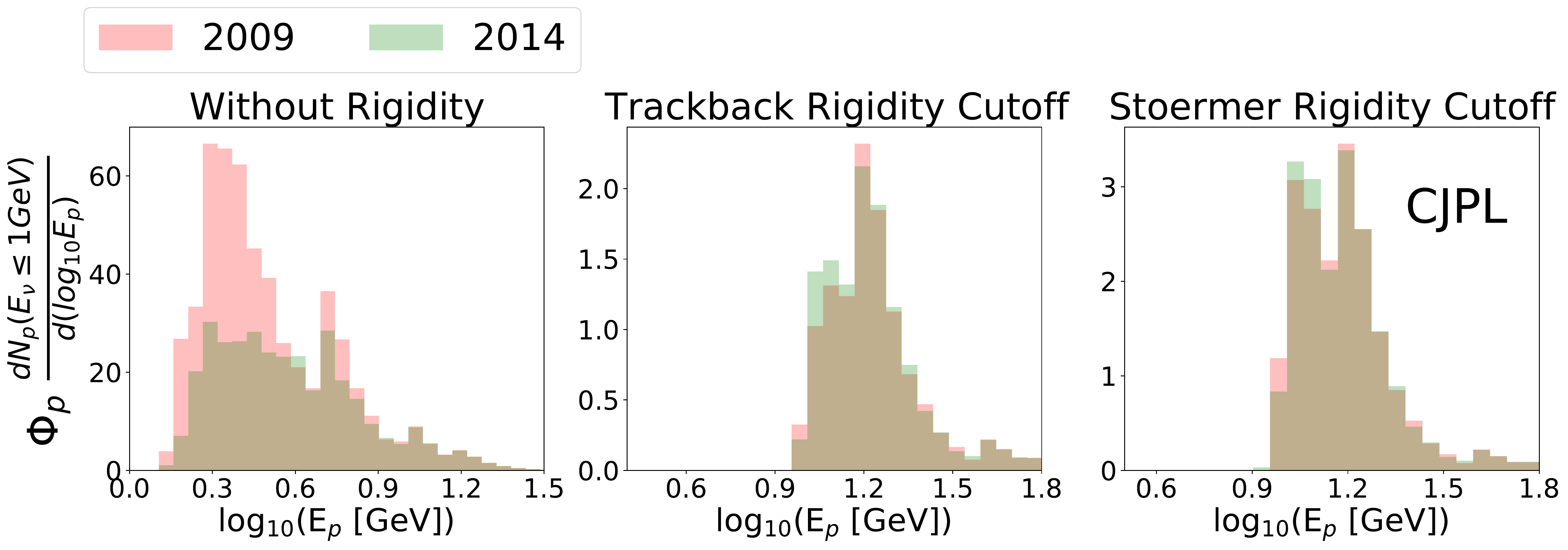}
    \includegraphics[width = 0.45\textwidth]{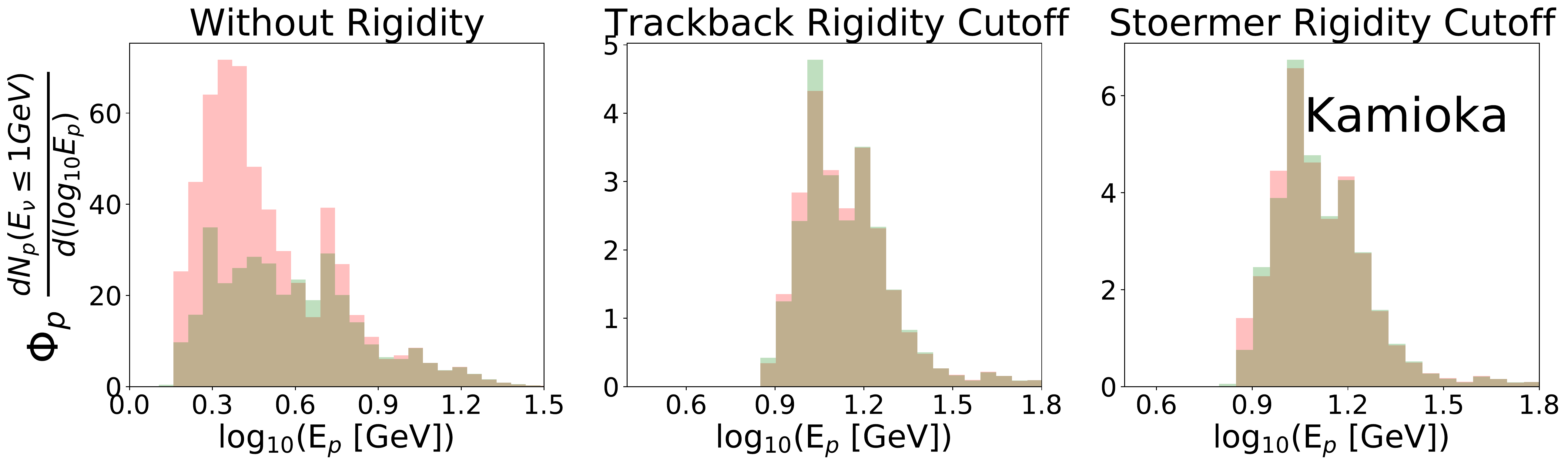}
    \includegraphics[width = 0.45\textwidth]{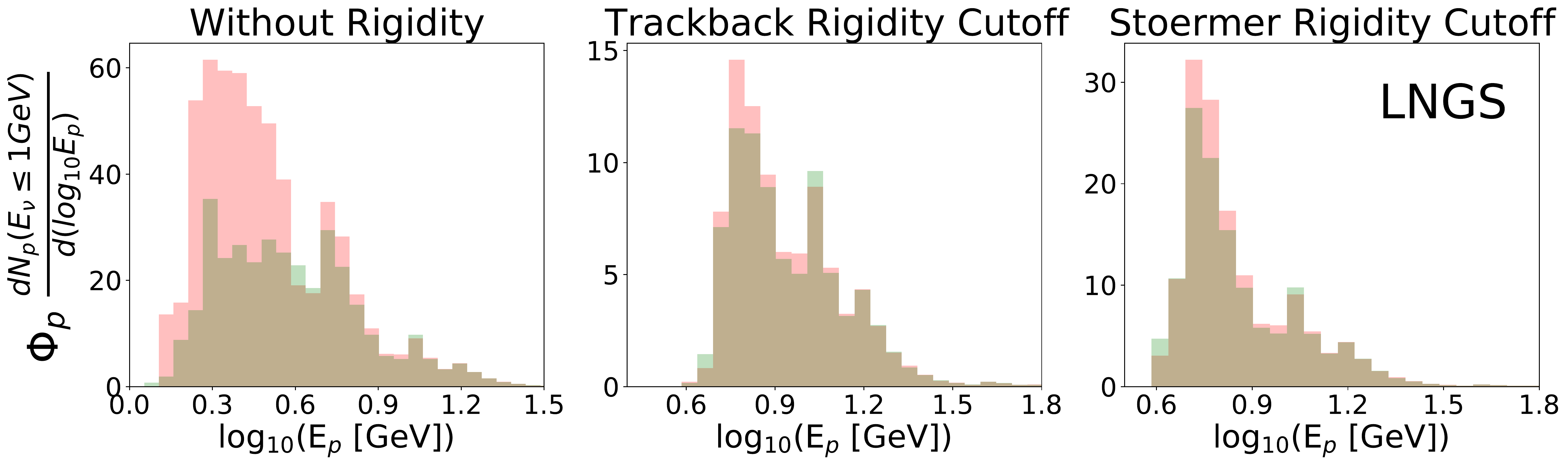}
    \includegraphics[width = 0.45\textwidth]{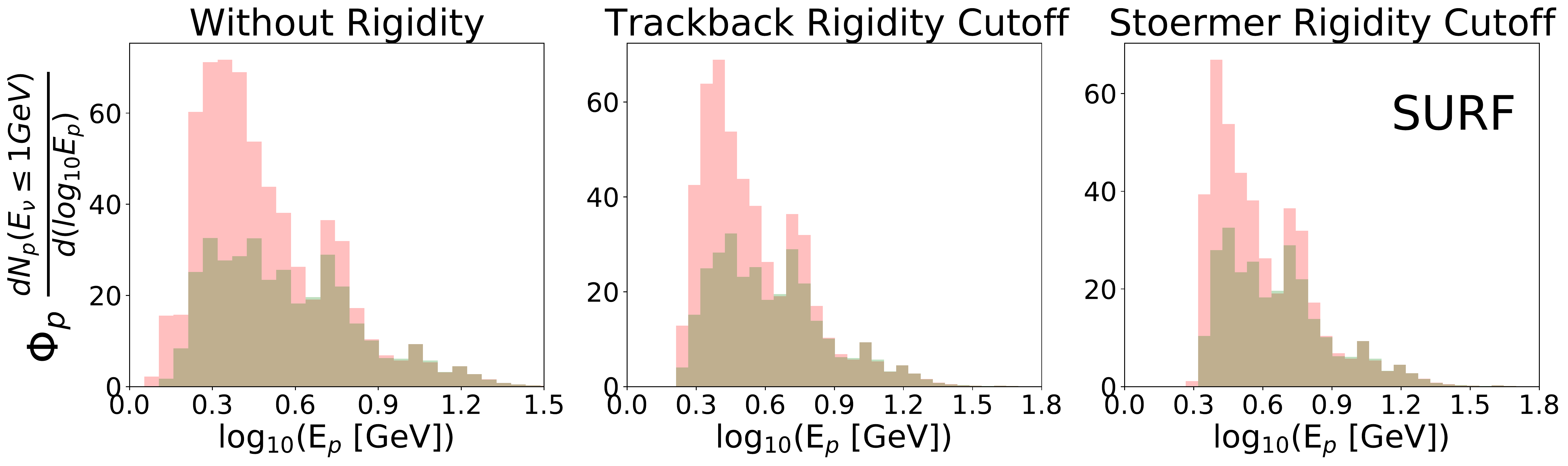}
    \includegraphics[width = 0.45\textwidth]{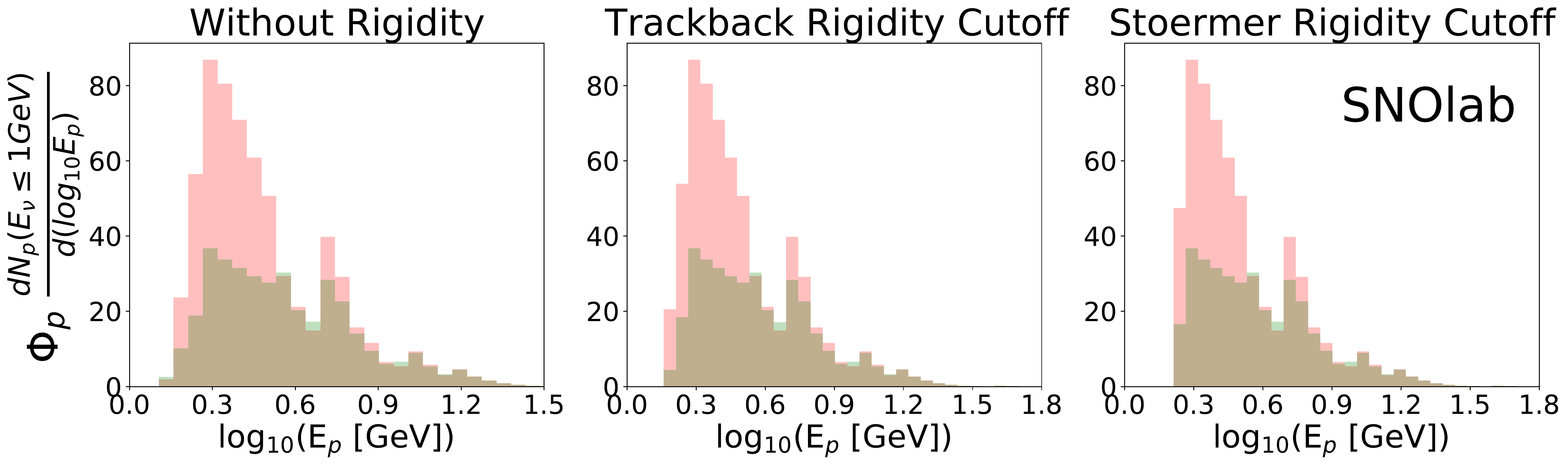}
    \caption{Histograms of cosmic ray (proton and helium) total energies that contribute at least one neutrino with energy less than or equal to 1 GeV at CJPL, Kamioka, LNGS, SURF and SNO. The first column is for the hypothetical case of no rigidity cut-off, the second column is for the trackback rigidity model, and the third column is for the Stoermer rigidity model. The normalizations and slopes differ between the years (see Equation~\ref{eqn: histperflux}). The peaks at 5 GeV and 10 GeV arise from weighting by different slopes and normalizations in each energy range.}
    \label{fig:protonhist}
\end{figure}

We are primarily interested in the neutrino flux less than approximately \mbox{$1$ GeV}, so we first identify the energy distribution of CRs from the CORSIKA simulations that produce neutrinos in this energy range. We calculate this energy range by defining the following: 
\begin{equation}
\begin{split}
        & \left [  \Phi_\mathrm{CR} \frac{dN_\mathrm{CR}(E_{\nu} \le 1\mathrm{GeV})}{d \ln(E_\mathrm{CR})}\right ] d \ln(E_\mathrm{CR}) \\
        = &\left [\Phi_\mathrm{CR} \frac{dN_\mathrm{CR}(E_{\nu} \le 1\mathrm{GeV})}{dE_\mathrm{CR}} \frac{dE_\mathrm{CR}}{d \ln(E_\mathrm{CR})} \right ] d \ln(E_\mathrm{CR}) \\
        = &\left [  \mathrm{Norm} \times E_\mathrm{CR}^{\gamma} \times \frac{dN_\mathrm{CR}(E_{\nu} \le 1\mathrm{GeV})}{dE_\mathrm{CR}} \times E_\mathrm{CR}\right ] d \ln(E_\mathrm{CR})\\
        = &\left [ \mathrm{Norm} \times E_\mathrm{CR}^{\gamma+1} \times \frac{dN_\mathrm{CR}(E_{\nu} \le 1\mathrm{GeV})}{dE_\mathrm{CR}}\right ] d \ln(E_\mathrm{CR}). 
\end{split}
\label{eqn: histperflux}
\end{equation}
In Figure~\ref{fig:protonhist}, we plot the quantity in brackets in Equation~\ref{eqn: histperflux}. This quantity is defined such when integrated gives the number of CRs that produce neutrinos with \mbox{$E_\nu~\le~1$ GeV} per primary proton or helium flux. If normalized by the total number of protons (or helium nuclei), Equation~\ref{eqn: histperflux} becomes the fraction of the primary proton (or helium) flux that contributes to the production of these neutrinos. 

In the histograms in Figure~\ref{fig:protonhist}, we examine a hypothetical model without a rigidity cut, and under the two assumed geomagnetic models discussed in the previous section. In addition, these are shown for both solar minimum and solar maximum. The sharp features at several energies arise from the different power law slope and normalization fits over the different energy ranges, as indicated in Table~\ref{tab: slope_norm}. 

Overall, these histograms show that protons with energy below \mbox{$\lesssim 10$ GeV} dominate the contribution to the neutrino flux \mbox{$\lesssim 1$ GeV} at all locations. In the panels without the rigidity cut-off, all detector locations show a reduced proton number as energies approach \mbox{$\lesssim 5$ GeV}. In the panels including the rigidity cut-offs, the proton distributions are more complex and more strongly depend on detector location. Comparing the different detector locations, we see that the  variation of proton number per primary proton flux increases between solar maximum and solar minimum at higher latitude, and they are more consistent at lower latitude.

\subsection{Modification to CORSIKA flux}

In order to most accurately estimate the atmospheric neutrino flux below \mbox{1 GeV}, we make modifications and additions to the output from CORSIKA. We now proceed to describe these, and how they are used to generate the flux predictions. 

\subsubsection{Zenith angle distributions}

We generate CR primaries isotropically in the half-steradian, so that their distribution is proportional to $\cos \theta_p$, where $\theta_p$ is the zenith angle of the incoming CR. For each neutrino energy bin, we choose $80$ bins in neutrino zenith angle, $\theta_\nu$, over the range \mbox{$-1 < \cos \theta_{\nu} <1$}. The choice of $80$ zenith bins is motivated by previous studies~\cite{2003Barr,2004Barr}, and corresponds to a width of \mbox{$\Delta \cos \theta_{\nu}$ = 0.025}. We find that this bin width optimizes in creating a smooth zenith angle distribution and avoiding large bin-to-bin fluctuations. In order to account for the projection of the detector area onto the solid angle within which the neutrinos are produced, we must weigh each angular bin by the factor $1/\cos \theta_{\nu}$, where $\theta_{\nu}$ is the median of the zenith angle of the bin. 

Once the data is binned in this way, we then renormalize the weighted angular flux to match the flux from the full HKKM simulation at \mbox{$\cos \theta_{\nu} = 0.5$}. To account for the different energy and angular binning between our simulations and that of HKKM, we compare the zenith distribution at each energy bin center with that of the closest HKKM energy. The renormalization for each energy bin center is then the average of matching our bin centers of \mbox{$\cos \theta_{\nu} = 0.4625, 0.5625$} with the corresponding centers at \mbox{$\cos \theta_{\nu} = 0.45, 0.55$} in the HKKM data, so each energy is associated with a different renormalization factor. This renormalization to match HKKM after $1/\cos\theta_{\nu}$ weighting is necessary because the angular distribution at certain energies depends on the energy bin size, interaction model and the angular bin size. We use \mbox{$\cos \theta_{\nu} = 0.5$} because it appropriately weights both the large and small zenith angle flux components (see also Ref.~\cite{2016Schoneberg}); the weighting factors decrease from 80 to 1 as $\cos \theta_{\nu} \rightarrow 1$. So the zenith distribution is enhanced more towards the horizontal and remains unchanged towards the vertical, with $0.5$ is sitting between these extreme cases. We average the values for energy bins down to \mbox{100 MeV} to reduce bin-by-bin fluctuations, so then the modification to the downward neutrino flux is a weighting of each angular bin and multiplication by an overall a renormalization constant, which varies in the range $\sim 20-40\%$ for the range of energy bins that we consider.

\begin{figure}[!htbp]
    \includegraphics[width = 0.5 \textwidth]{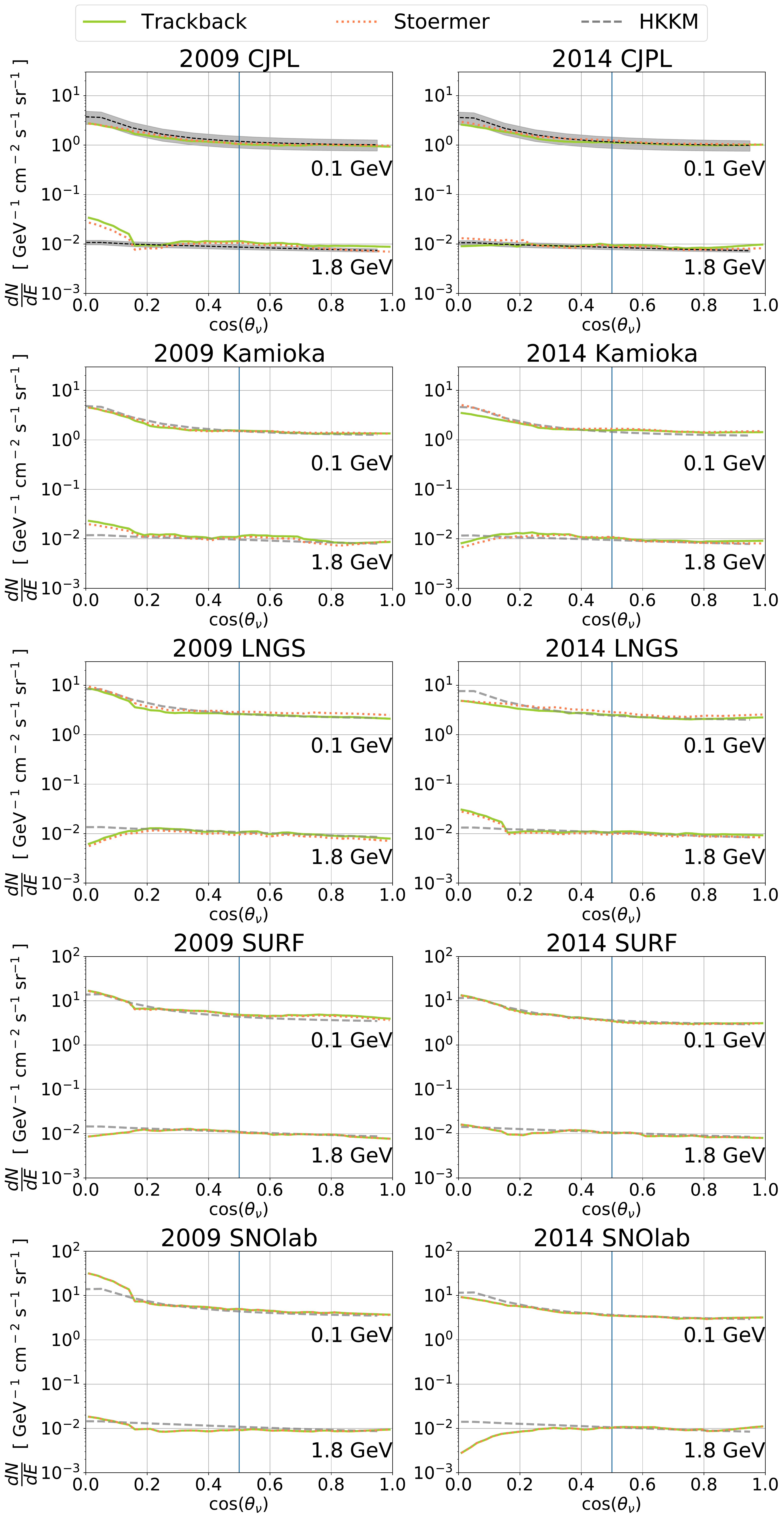}
    
    \caption{Zenith angle distribution at \mbox{100 MeV} and \mbox{1.8 GeV} from our CORSIKA simulations as compared to the full-3D HKKM results (dashed). For our simulations, shown are the results for the Stoermer (red) and trackback (green) cut-off models. Left column is for 2009, and right column is for 2014. Each of the five rows gives a different detector location. For CJPL, the grey shade is the range encompassing Kamioka and INO and the dashed black line is their average.}
    \label{fig:zenith}
\end{figure}

The above matching is performed at all detector locations; we note that for CJPL, because of its location, the HKKM angular distribution is the average of that at Kamioka and the India-based Neutrino Observatory (INO)~\citep{ICAL:2015stm}. For all locations, our choice of zenith angle binning produces a smooth zenith distribution that agrees with the shape of that in HKKM (Fig. \ref{fig:zenith}). The inconsistent features toward the horizon comes from numerical fluctuations. Note that, as we show below, our angle-integrated flux  matches the HKKM normalization as well as the measured SK electron neutrino flux. 

After this weighting and renormalization, the flux is then integrated over the half-steradian $2\pi$ to get the downward neutrino flux as a function of energy. The integrated downward flux is then smoothed by a Savitzky-Golay Filter using Scipy package~\cite{2020SciPy-NMeth}, with window length \mbox{N = 5} and polyorder \mbox{M= 3}. These values of N and M are chosen based on the criteria \mbox{N $<$ 2M}, and to optimize the smoothing of the flux at both low and high energies.

\subsubsection{Up-down ratio}

At any location, the neutrino flux comes from a downward ($0 < \cos \theta_{\nu} < 1$) and an upward component ($-1 < \cos \theta_{\nu} < 0$). We simulate the downward flux locally at the detector as described above. Because the upward flux requires knowing the rigidity cut-off at all positions for all directions, determining it presents a substantial computational challenge. An additional complexity is that only a small fraction of neutrinos from any given direction pass through the detector~\citep{2003Wentz,Honda:2006qj}. 

To obtain the upward-going flux, we make several simplifying though well-motivated assumptions. First, the ratio of the upward-to-downward flux across all energies is nearly unchanged in going from the 1D simulation to the full 3D simulation~\cite{Barr:2004br}. Therefore for computational efficiency, we simulate CORSIKA events for down-going zenith angles as above, and use the previously-determined up-down ratios to obtain the upward flux. 

An additional complication arises because the up-down ratio is not constant across the solar cycle. HKKM~\cite{Honda:2015fha} have determined the time-dependent up-down ratio for neutrino energies \mbox{$> 100$ MeV}, while Ref.~\cite{2002updown} extends down to lower energies of \mbox{$50$ MeV}, though the latter authors do not include time dependence. 

In our calculations, to obtain the upward flux, we scale our downward-going fluxes from CORISKA by the up-down ratio for each detector location. For Kamioka, LNGS and SNO, we use HKKM data down to \mbox{$100$ MeV}, and the ratio from Ref.~\cite{2002updown} below \mbox{$100$ MeV}. For SURF, we use only HKKM data down to \mbox{$100$ MeV} since there is no available data for lower energies. For CJPL, since there is no published up-down ratio, we use the up-down ratio as the average of that from the HKKM fitting at INO and Kamioka. Note that we carry the uncertainties in the up-down ratio through into the calculations of the events rates at the detectors in the sections below. 

\begin{figure}[!htbp]
    \includegraphics[width = 0.4\textwidth]{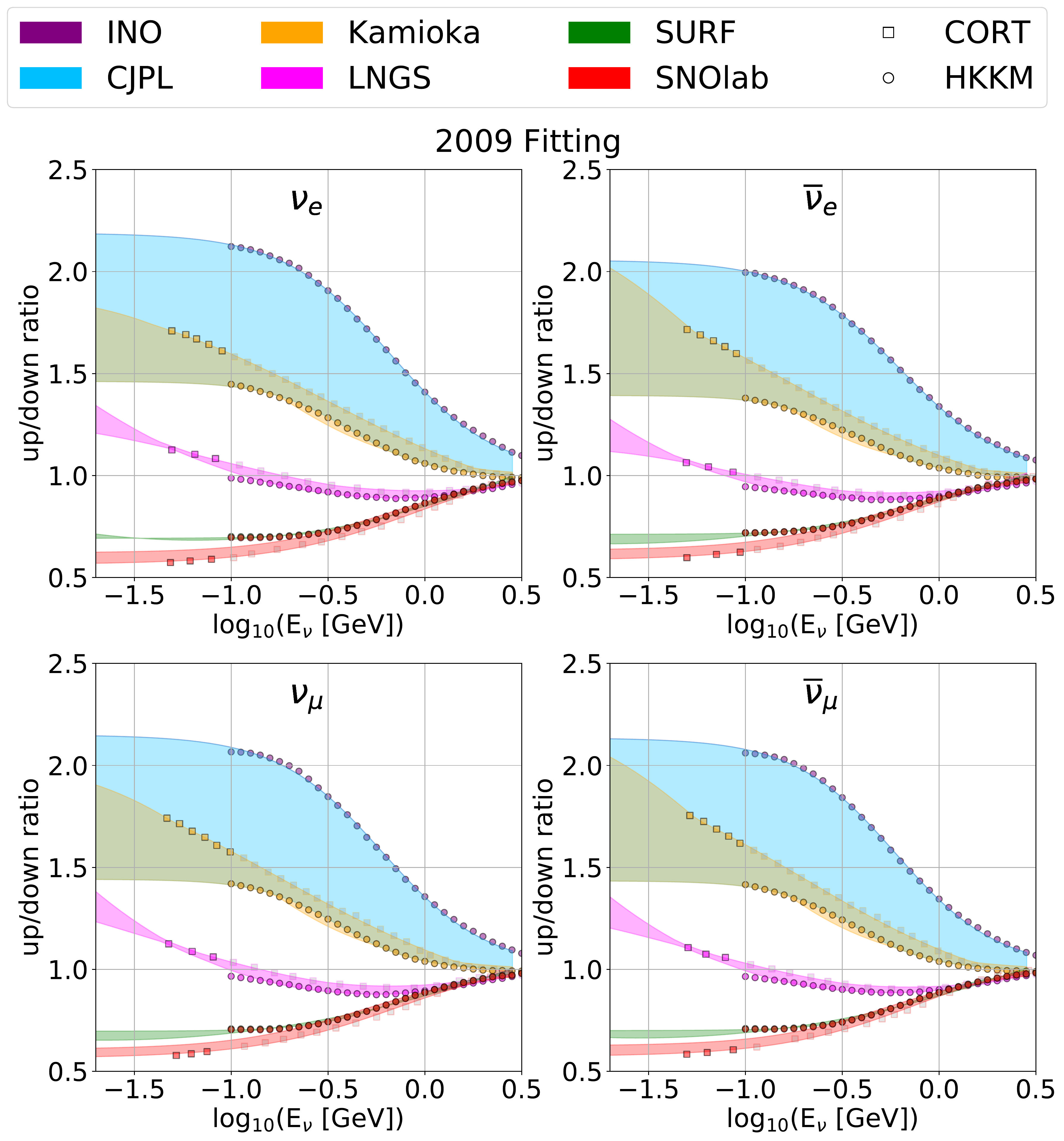}
    \includegraphics[width = 0.4\textwidth]{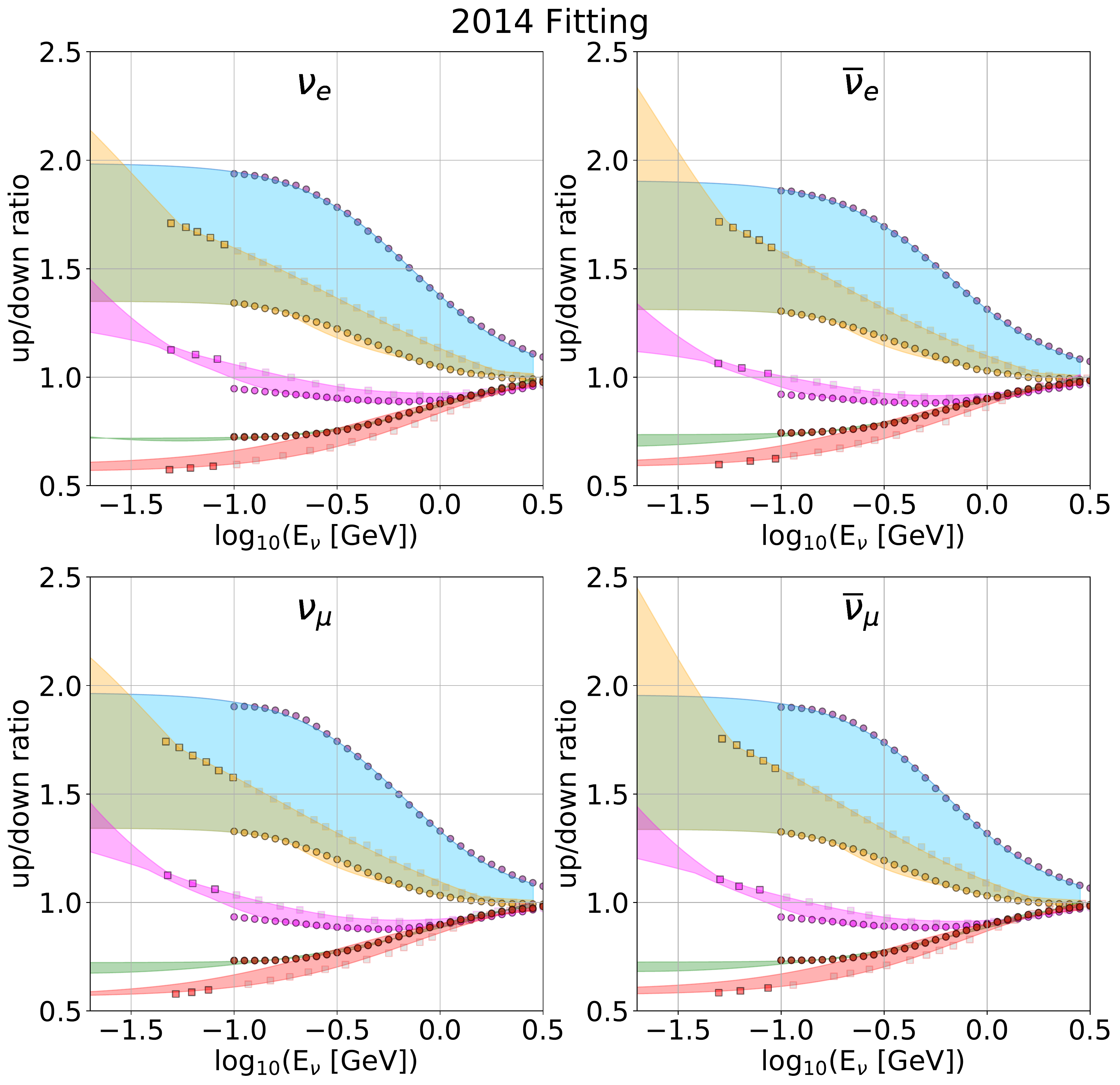}

    \caption{Up-down ratios from HKKM~\citep{Honda:2015fha,hondadata} and from the CORT~\cite{2002updown} simulations. Top two rows are for 2009, and bottom two rows are for 2014. Each panel is for a different neutrino flavor as indicated.}
    \label{fig:updownratio}
\end{figure}

The error bands for the up-down ratios that we use for each detector location are shown in Figure~\ref{fig:updownratio}. Since we calculate the flux to energies below where measurements of the up-down ratio have been measured, we conservatively define error bands as indicated to bracket reasonable boundaries for the extrapolated up-down ratio to lower energies. This shows that for our calculation the uncertainty in  up-down ratio is largest at Kamioka and CJPL, and is negligible in higher latitude locations.

\subsubsection{Neutrinos from stopped muons at sea level and pions, muons decaying at rest}

CORSIKA traces hadrons and muons with kinetic energy (ECUT) down to \mbox{$20$ MeV} and \mbox{$10$ MeV}, respectively. As discussed below, this produces a systematic uncertainty in some of our rate predictions that must be accounted for. Part of this very low-energy neutrino flux that is below the CORISKA threshold is due to stopped muons that decay or capture after hitting the surface of the Earth. These muons contribute a neutrino flux component with an energy spectrum similar to the shape of neutrinos from muon decay at rest. 

When high energy muons produced from pions reach the surface of the Earth, $\mu^{+}$ simply decay via  
\begin{equation}
\begin{split}
    \pi^{+} \longrightarrow & \mu^{+} + \nu_{\mu}\\
   & \mu^{+} \longrightarrow  \mathrm{e}^{+} + \nu_ \mathrm{e} + \bar \nu_{\mu}
\end{split}
    \label{eqn: muonplus_decay}
\end{equation}
while $\mu^{-}$ either decay or are captured, i.e. 
\begin{equation}
    \begin{split}
    \pi^{-} \longrightarrow & \mu^{-}+\bar \nu_{\mu}\\
        &\mu^{-} + N_\mathrm{1}  \longrightarrow N_\mathrm{2} + \nu_{\mu}\\
        &\mu^{-} \longrightarrow \mathrm{e}^{-} + \bar \nu_\mathrm{e} + \nu_{\mu}
    \end{split}
    \label{eqn: muonminus_decay}
\end{equation}
where $N$ is a nucleus. The flux from this component can be calibrated to the muon flux at sea level. The flux from this component is~\cite{2019stoppingmuon}
\begin{equation}
    \phi_\mathrm{rest} = f_{\nu}J_{\mu_{\pm}}\frac{R_{\oplus}}{4(R_{\oplus}-d)}\ln{\frac{R_{\oplus}^{2}+(R_{\oplus}-d)^2+2R_{\oplus}(R_{\oplus}-d)}{R_{\oplus}^{2}+(R_{\oplus}-d)^2-2R_{\oplus}(R_{\oplus}-d)}}
    \label{eqn: decay_at_rest}
\end{equation}
where $d$ is the detector depth, $f_{\nu}$ is neutrino spectrum from stopped muon decay per flavor, and $J_{\mu_{\pm}}$ is the muon flux integrated over momentum and solid angle. To perform this calculation, we practically assume \mbox{$d = 0.1$ m} to avoid a divergence since the detectors are at sea level, and we obtain the muon number arriving at sea level from our CORSIKA simulations. For comparison, \mbox{$\phi_\mathrm{rest} = 6.62f_{\nu}J_{\mu_{\pm}}$} using the approximation in Ref.~\cite{2019stoppingmuon}. The difference in total flux between these different normalizations of $\phi_\mathrm{rest}$ is $\sim$ \mbox{7$\%$}. From our simulations, we confirm that the variation of the intensity of the positive and negative muon components between solar maximum and solar minimum, $J_{\mu^{+}}$/$J_{\mu^{-}}$, is consistent with BESS results~\citep{2004BESS}. We take the percentage of $\mu^{-}$ undergoing a decay process as 60.65$\%$, and the $\nu_{\mu}$ spectrum from $\mu^{-}$ nuclei capture spectrum is adopted from the photon spectrum reaction with $^{16}$O \cite{1979Strassner}. This latter spectrum terminates at \mbox{40 MeV}, so we can ignore it because of its small contribution relative to the decay at rest $\nu_{\mu}$ spectrum. To test the fluctuation in total neutrino flux for a deeper detector such as SNO, we vary the detector depth to \mbox{2100 m}. The change in total flux for \mbox{$E_\nu < 53$ MeV} between detector depth \mbox{$0.1$ m} and \mbox{$2100$ m} is only $\lesssim$ 5$\%$, using either Equation~\ref{eqn: decay_at_rest} or the approximate equation in Ref.~\cite{2019stoppingmuon}. 

\subsection{Resulting Flux}
\begin{figure}[!htbp]
    \includegraphics[width = 0.4\textwidth]{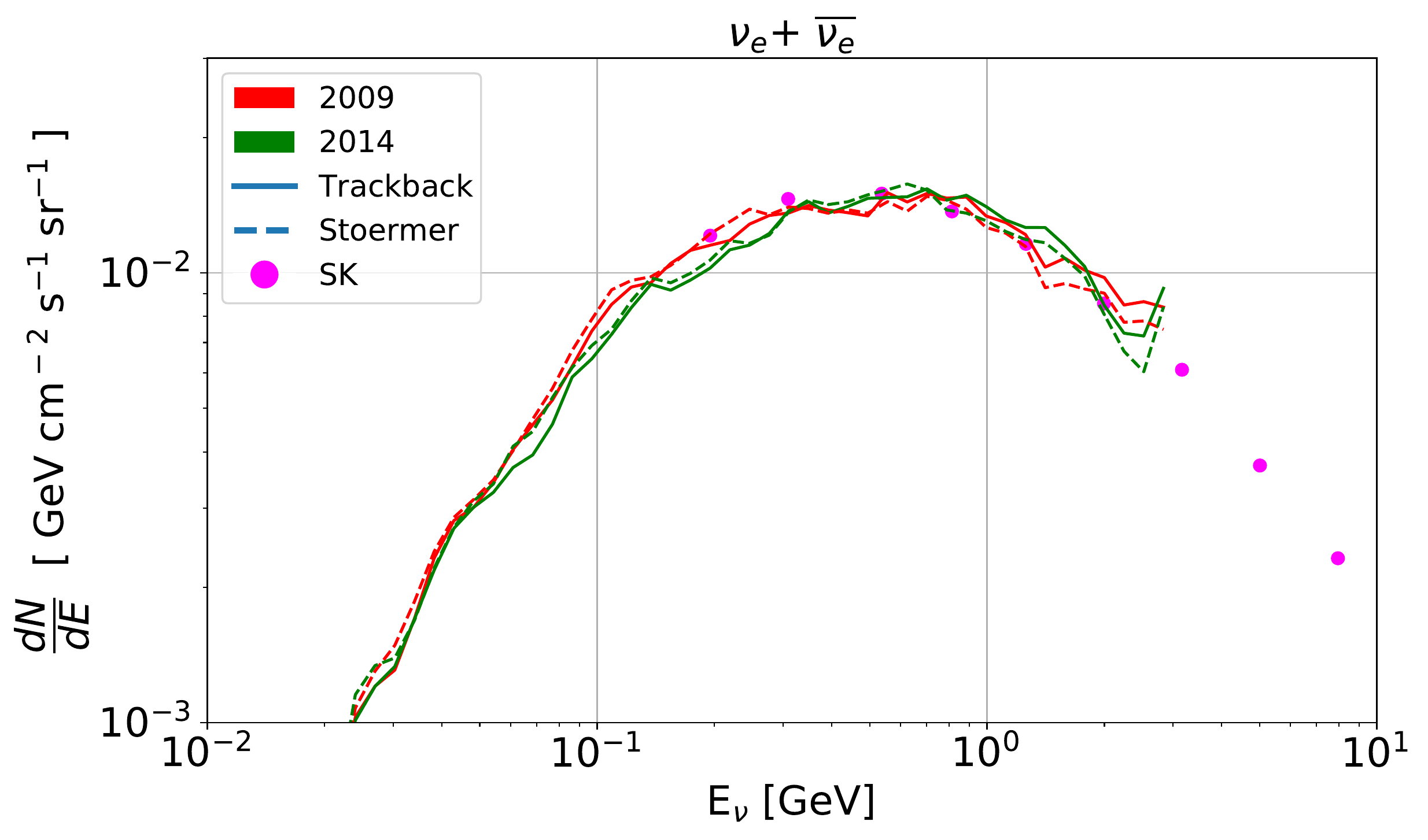}
    \caption{Electron neutrino flux at Kamioka, from our simulations in 2009 and 2014, for the Stoermer and trackback models. The Super-Kamiokande results for the flux are shown as the data points.}
    \label{fig:SPK_REAL_nue}
\end{figure}

Combining all of the components above, we estimate the atmospheric neutrino flux for all flavors down to energies below \mbox{$\lesssim 100$ MeV} for all of our detector locations. To calibrate our calculation to that of existing data at higher energies, we compare to the electron neutrino flux at Kamioka in Figure~\ref{fig:SPK_REAL_nue}. We use the electron neutrino flux for comparison because this component is unaffected by neutrino oscillations at these energies and path lengths. Shown is the flux calculated at solar minimum and at solar maximum. The measured flux by SK phases I-IV covers a time span from 1996 to 2016. Note that Kamioka is less affected by solar modulation because of its high-rigidity cutoff. 

\begin{figure*}[!htbp]
    \includegraphics[width=0.8\textwidth]{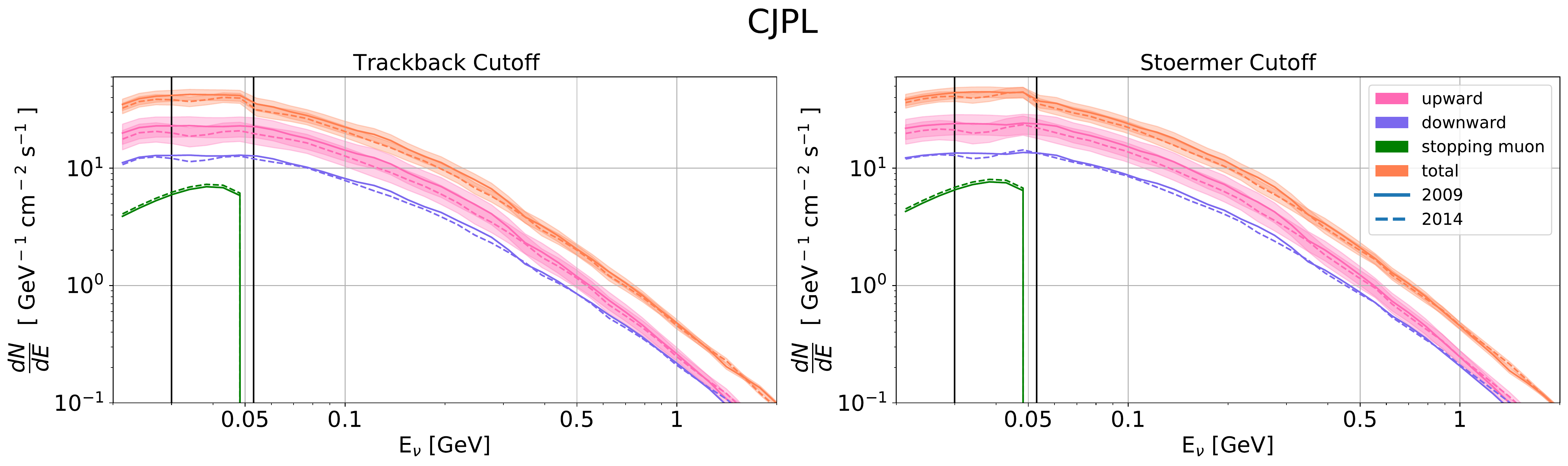}
    \includegraphics[width=0.8\textwidth]{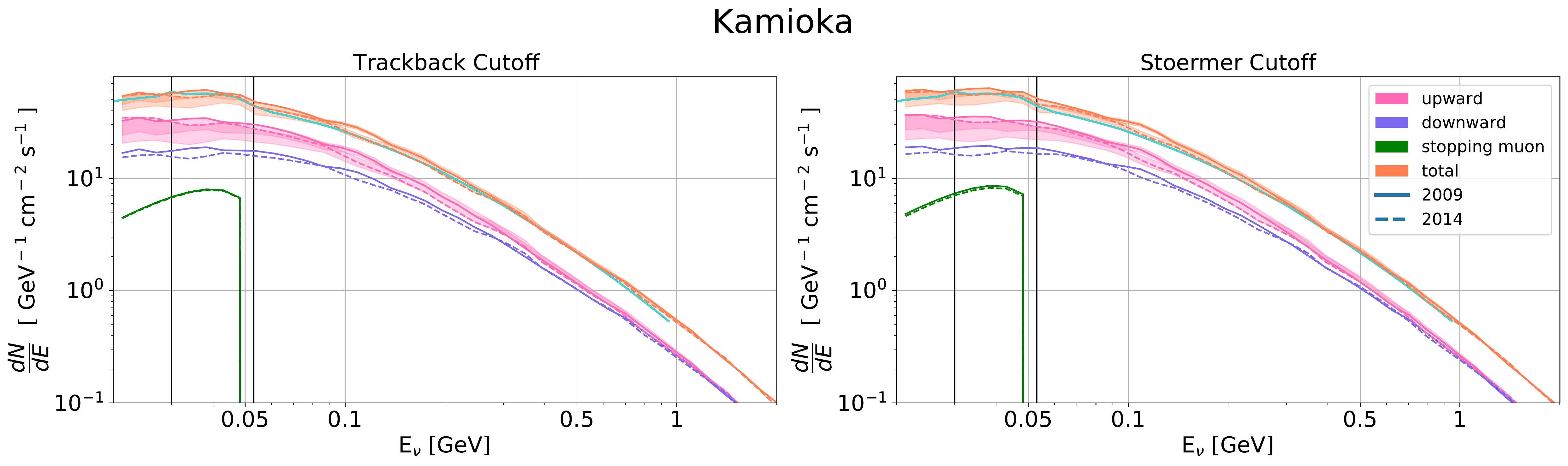}
    \includegraphics[width=0.8\textwidth]{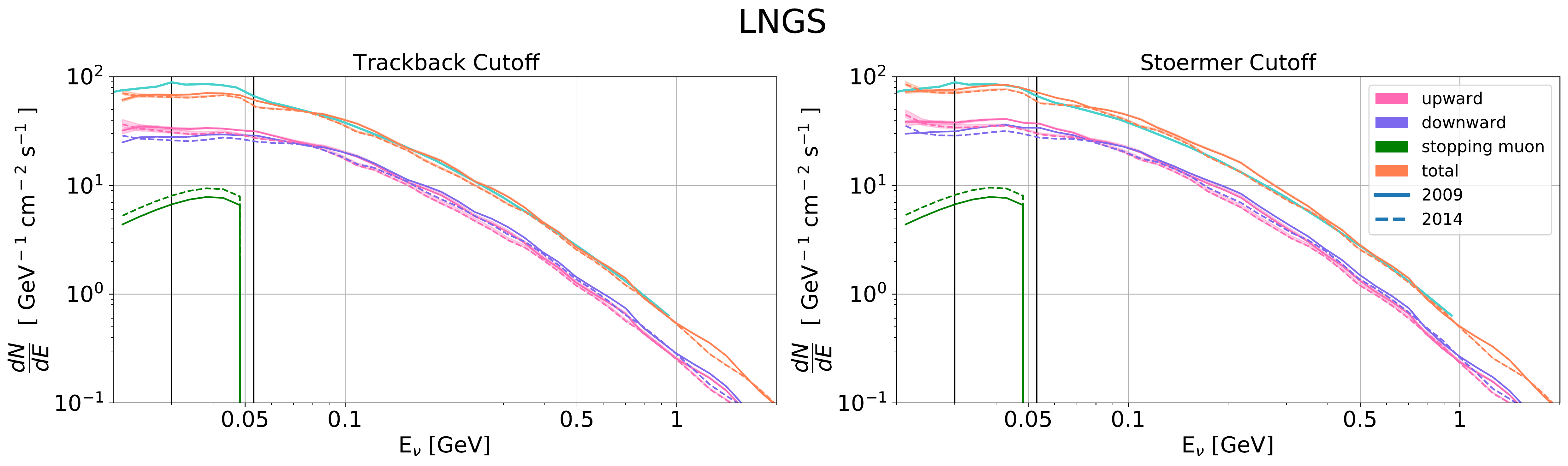}
    \includegraphics[width=0.8\textwidth]{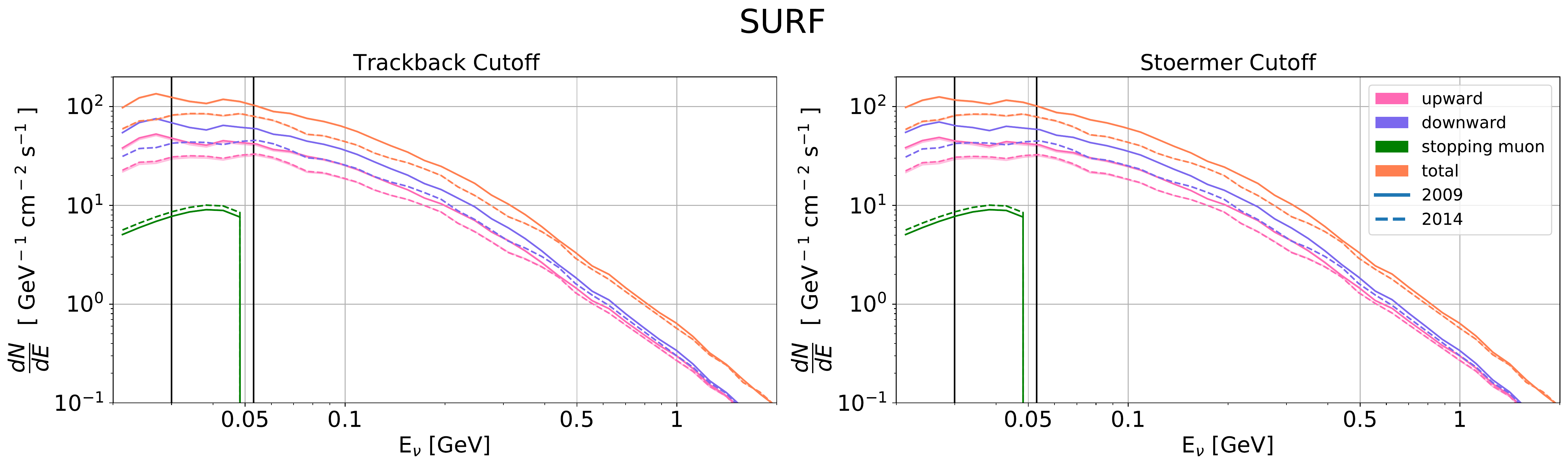}
    \includegraphics[width=0.8\textwidth]{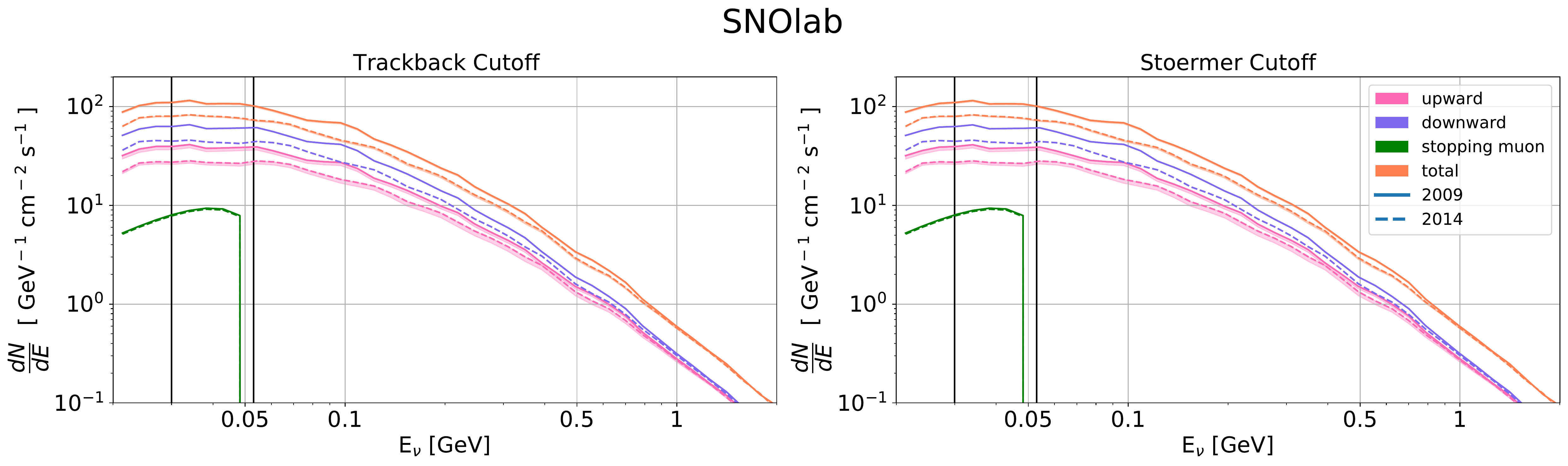}

    \caption{Neutrino fluxes for 2014 (solar maximum) and for 2009 (solar minimum). Left column is for the trackback model, and right column is for the Stoermer cut-off model. Each row is for a different detector location. For Kamioka and LNGS, we compare with the solar average results from FLUKA (cyan lines). The shaded regions reflect the errors from the assumed up-down ratio.}
    \label{fig:error_flux}
\end{figure*}

The flux predictions at all detector locations are shown in Figure~\ref{fig:error_flux}. Comparing with HKKM, at CJPL, Kamioka, LNGS, SURF, and SNO, the resulting downward flux matches the HKKM downward flux and the upward flux by using the up-down ratio also matches the HKKM upward flux, except for a slight under predictions at low energies \mbox{$\sim 100$ MeV}. Similarly the total flux matches the total flux in HKKM. 

In addition to the comparison to SK and HKKM, we can compare our results to the previous results from FLUKA~\citep{20003DBattistoni}, which are the only calculations that extend down to the neutrino energies considered here. The comparison with FLUKA shows that the neutrino flux from the trackback model is in better agreement with the FLUKA results than that from the Stoermer model is. Assuming the trackback model, the flux at Kamioka and the flux above \mbox{$100$ MeV} at LNGS at solar minimum 2009 are slightly higher than the solar-averaged FLUKA flux, and are slightly lower in solar maximum 2014. We find that the Stoermer model over predicts both the flux at solar minimum and solar maximum relative to FLUKA. Muons decaying at rest after reaching sea level increase the flux in the decay-at-rest regime \mbox{$E_\nu < 53$ MeV}. For LNGS, the flux decreases below \mbox{$100$ MeV} and we underestimate the flux below \mbox{$53$ MeV} in the trackback model by $\sim 25\%$ compared with FLUKA.   

Even with our modifications, the very low-energy neutrino flux ($E_{\nu}<53$ MeV) may still be  systematically lower than the true flux, for two reasons. First, because of the lower energy cut used in CORSIKA (ECUT), hadrons with kinetic energy less than \mbox{$20$ MeV} and muons with kinetic energy less than \mbox{$10$ MeV} are discarded. Examination of the weighted histogram of neutrinos from muon or pion decay shows that, in particular at high latitude locations SNO and SURF, there is a sharp decrease around 50 and \mbox{$30$ MeV}, indicating missing neutrinos from $\pi$ and $\mu$ decaying at rest. The effect is less severe at lower latitude locations which are less sensitive to low energy CRs. 

The second reason that our very low energy flux is likely an underestimate of the true flux is due to geomagnetic effects. As discussed above, high angular resolution is required to resolve the diffusive proton flux in the penumbra. Although the azimuthal asymmetry is achieved by both the Stoermer and the trackback model, some protons below the hard cut are missing at LNGS. At SNO and SURF, the proton flux does diffuse to lower energy in the trackback model but this has negligible affect on the neutrino flux.

Systematic errors in our flux calculation may also be incurred from our renormalization of the zenith angle distribution, and our use of the up-down ratio to scale and obtain the total flux. This is likely the most significant for SNO and SURF, with the scaling of the up-down ratio error dominating at Kamioka and CJPL. At LNGS, the neutrino flux is more significantly affected by the proton flux which results from the assumed geomagnetic model. Though our low-energy flux is likely systematically underestimated, we show below that our flux predictions are relatively unaffected for the majority of detector targets that we consider. 

\section{Coherent Elastic Neutrino-Nucleus Scattering \label{sec: scattering}}
Atmospheric neutrinos will interact in detectors we consider through the coherent elastic neutrino-nucleus scattering (CE$\nu$NS) process. This scattering proceeds through the exchange of a $Z$-boson within a neutral current interaction. The resulting differential neutrino-nucleus cross section as a function of the nuclear recoil energy $E_r$ and the incoming neutrino energy $E_\nu$ is~\cite{1977Freedman}
\begin{equation}
    \frac{d\sigma (E_{r}, E_{\nu})}{dE_{r}} = \frac{G_\mathrm{F}^2}{4\pi} Q_\mathrm{w}^2 M_\mathrm{target} \left ( 1 - \frac{M_\mathrm{target} E_{r}}{2E_{\nu}^2} \right ) F^{2}(E_{r}).
    \label{eqn: dsigmadT}
\end{equation}
The recoil energy of the target nuclei is related to neutrino energy by \mbox{$E_\mathrm{\nu,min}= \sqrt{\frac{M_\mathrm{target} E_{r}}{2}}$}. The weak nuclear charge is \mbox{$Q_\mathrm{w}= N - (1-4\sin^{2}\theta_\mathrm{w})Z$}, where \mbox{$\sin^{2}\theta_\mathrm{w} = 0.2223$}~\cite{2016codata} is the Weinberg angle. The mass of the target nucleus is \mbox{$M_\mathrm{target} = N M_\mathrm{n}+ZM_\mathrm{p}$}, where $N$, $Z$ are the number of neutrons and protons. The nuclear form factor is $F(E_r)$, which in part determines the loss of coherence in the scattering. For our analysis we use the Helm form factor~\cite{1956Helm}.  Nuclear effects may be discernible in CE$\nu$NS experiments, though we do not consider these in our analysis~\citep{Hoferichter:2020osn}. 

In the analysis below we consider xenon, argon, and helium targets, as these provide a plausible range of the types of nuclear targets being developed. With this range of target nuclei, it also allows us to study the phenomenology over a wide range of nuclear mass.

Integrating over neutrino energy subject to the kinematic limit gives the event rate off of a given target, 
\begin{equation}
    \frac{dR(E_{r})}{dE_{r}} = \int_{E_\mathrm{\nu,min}}^{\infty}  \Phi_{\nu}(E_{\nu}) \frac{d\sigma (E_{r}, E_{\nu})}{dE_{r}} dE_{\nu}, 
    \label{eqn: eventrate_Enu}
\end{equation}
where $\phi_\nu$ is the neutrino flux. For a given $E_\nu$ there is a corresponding maximum recoil energy and integrating over recoil energies with $E_r < \frac{2 E{_\nu}^{2}}{M_\mathrm{target}}$ gives the distribution of neutrino energies that a given target is sensitive to, 
\begin{equation}
    \frac{dR(E_{\nu})}{dE_{\nu}} = \Phi_{\nu}(E_{\nu}) \int_{E_\mathrm{r,min}}^{E_\mathrm{r,max}}  \frac{d\sigma (E_{r}, E_{\nu})}{dE_{r}} dE_{r}
    \label{eqn:eventrate_Er}
\end{equation}

\begin{figure*}[!htbp]
    \centering

    \includegraphics[width =0.9\textwidth]{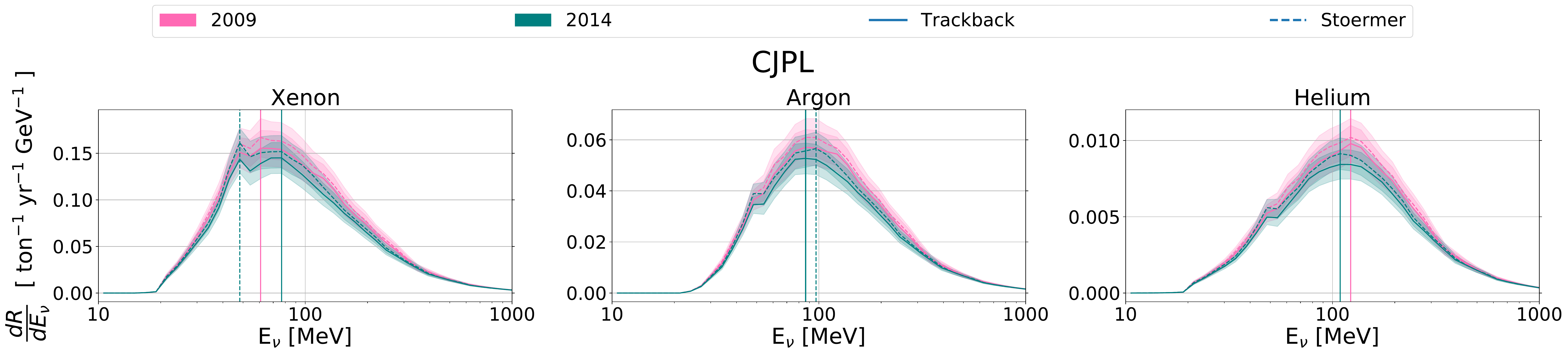}
    \includegraphics[width =0.9\textwidth]{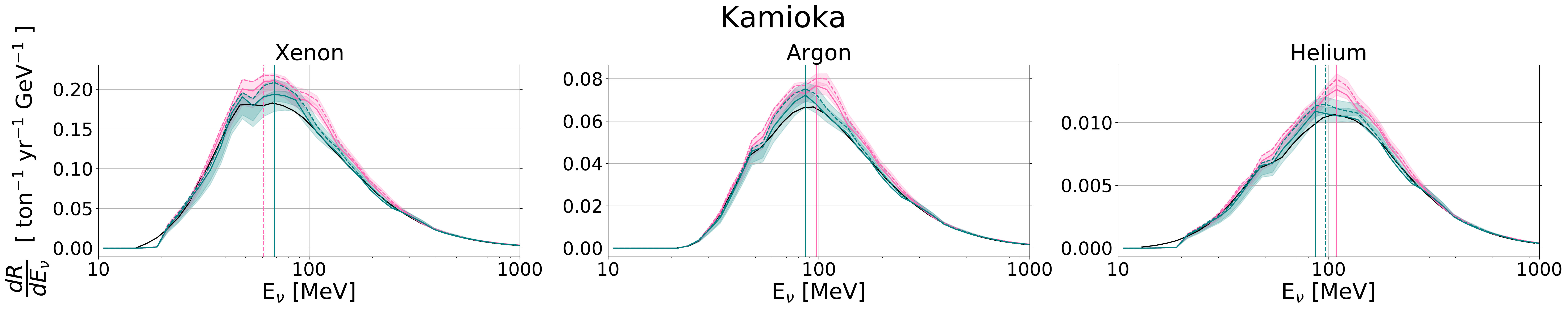}
    \includegraphics[width =0.9\textwidth]{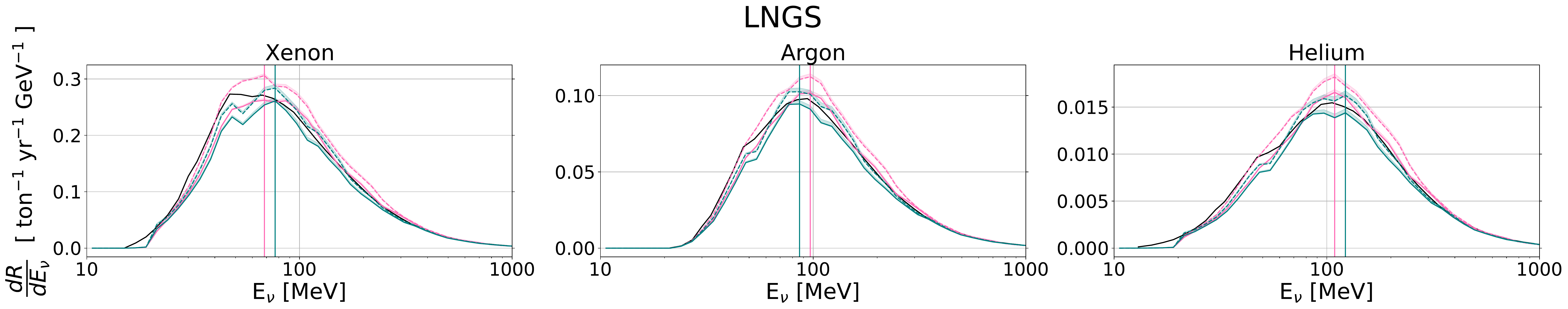}
    \includegraphics[width =0.9\textwidth]{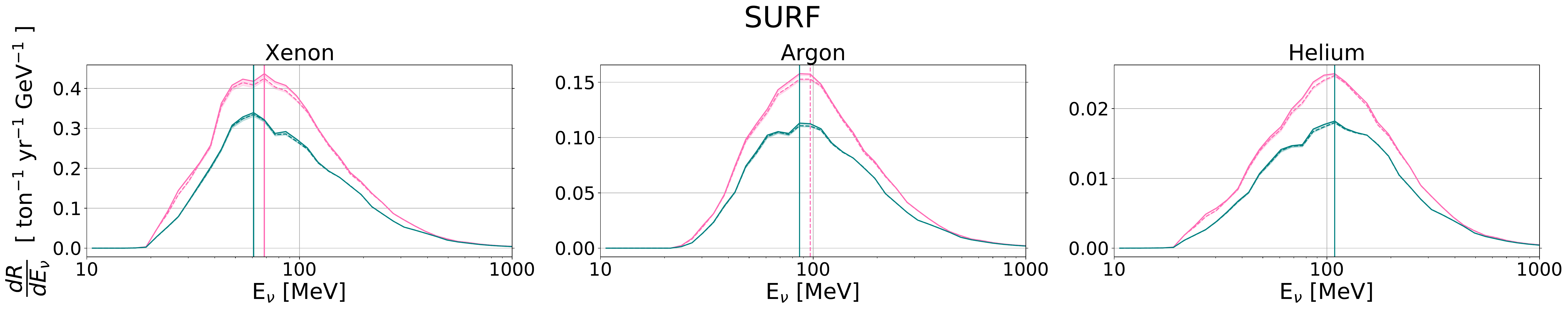}
    \includegraphics[width =0.9\textwidth]{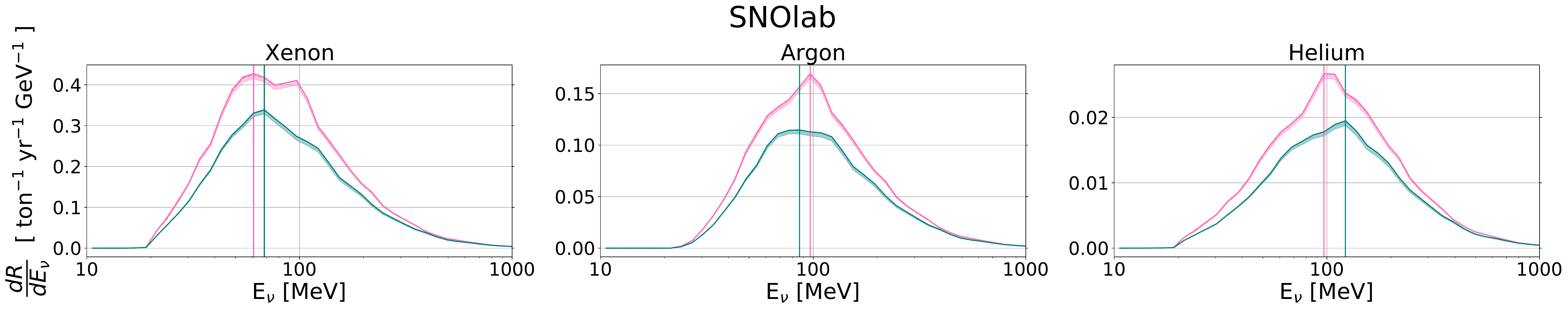}
    
    \caption{Distributions of neutrino energies that each detector target and each detector location are sensitive to. From left to right, the columns are for xenon, argon, and helium detectors, and the rows are for a different detector location. The distributions are shown for solar minimum and solar maximum, and for both the trackback and the Stoermer cut-off model. Vertical lines indicate the position of the peak of the distributions. Shaded bands represent the uncertainty due to the assumed up-down ratio. The black curves in the Kamioka and LNGS panels are the FLUKA solar average results.}
    \label{fig: rate_vs_Enu}
\end{figure*}

\begin{figure}[!htbp]
    \includegraphics[width = 0.5\textwidth]{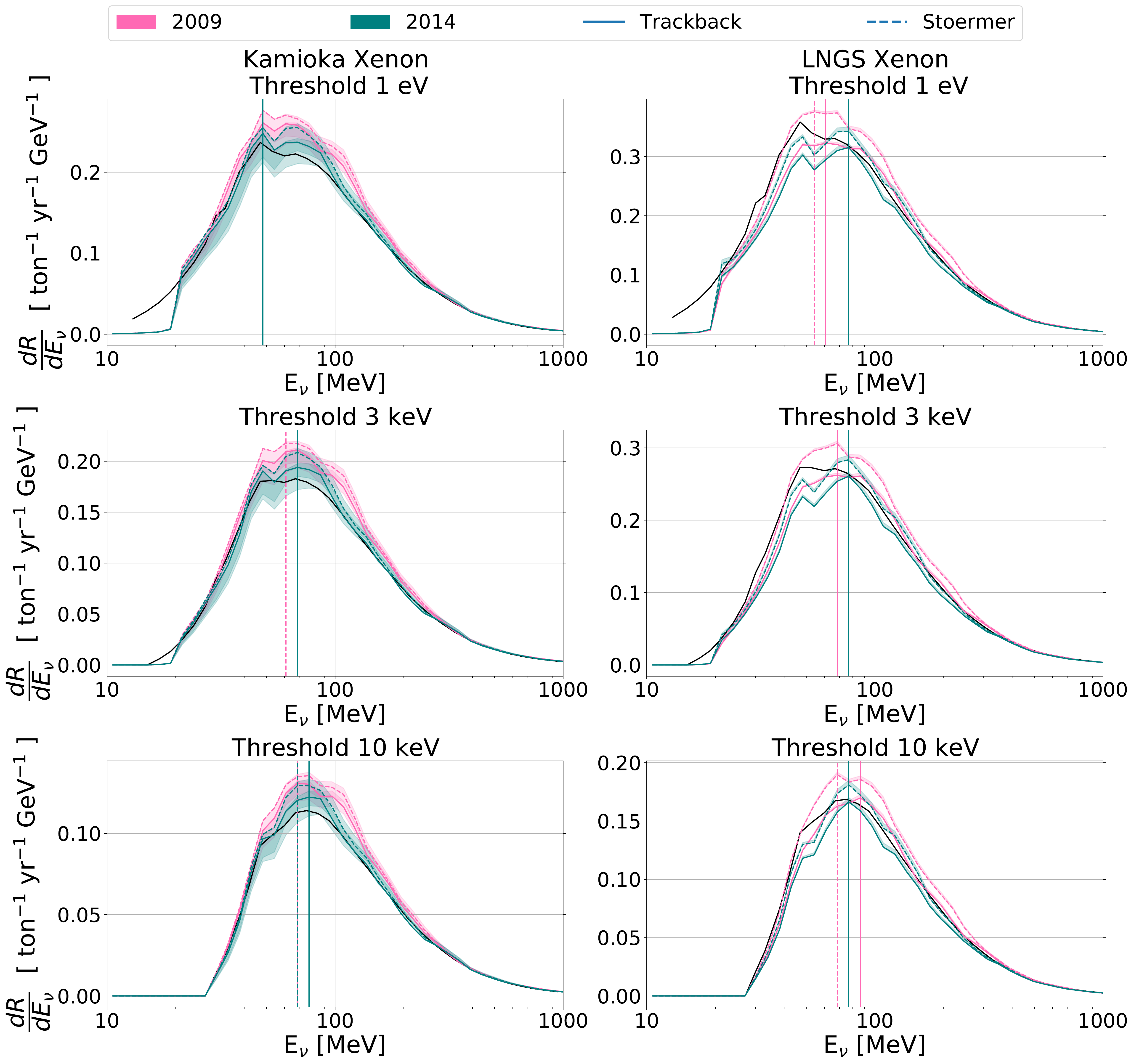}
    \caption{Energy distribution of atmospheric neutrinos observed in a xenon detector at Kamioka and LNGS, for various recoil energy thresholds. The distributions are shown for solar minimum and solar maximum, and for both the trackback and the Stoermer cut-off model. Vertical lines indicate the position of the peak of the distributions.  Shaded bands represent the uncertainty due to the assumed up-down ratio. The black curves are the FLUKA solar average results. }
    \label{fig: Xe_threshold}
\end{figure}

\begin{figure*}[!htbp]
    \centering
    \includegraphics[width=0.8\textwidth]{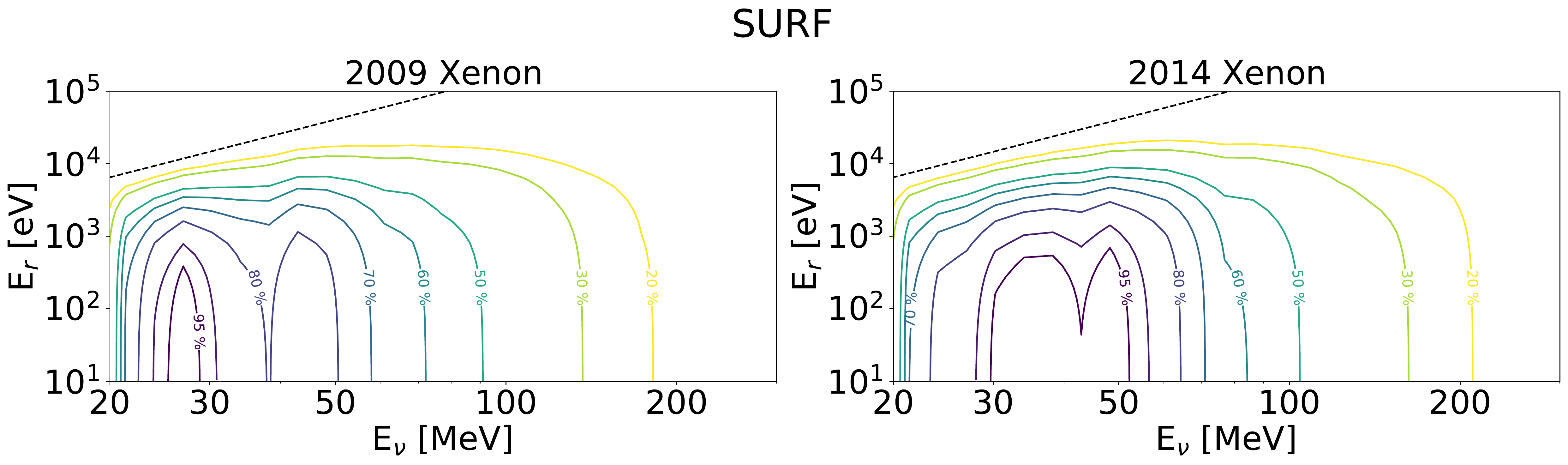}
    \caption{Normalized detector response function of xenon at SURF. Shown are the results for solar minimum (2009) and solar maximum (2014) for trackback geomagnetic model. Contour lines are percentage with respect to the maximum. The lines indicate the kinematic limit.}
    \label{fig:trackbackxenon}
\end{figure*}

\begin{figure*}[!htbp]
    \centering
    \includegraphics[width = 0.9\textwidth]{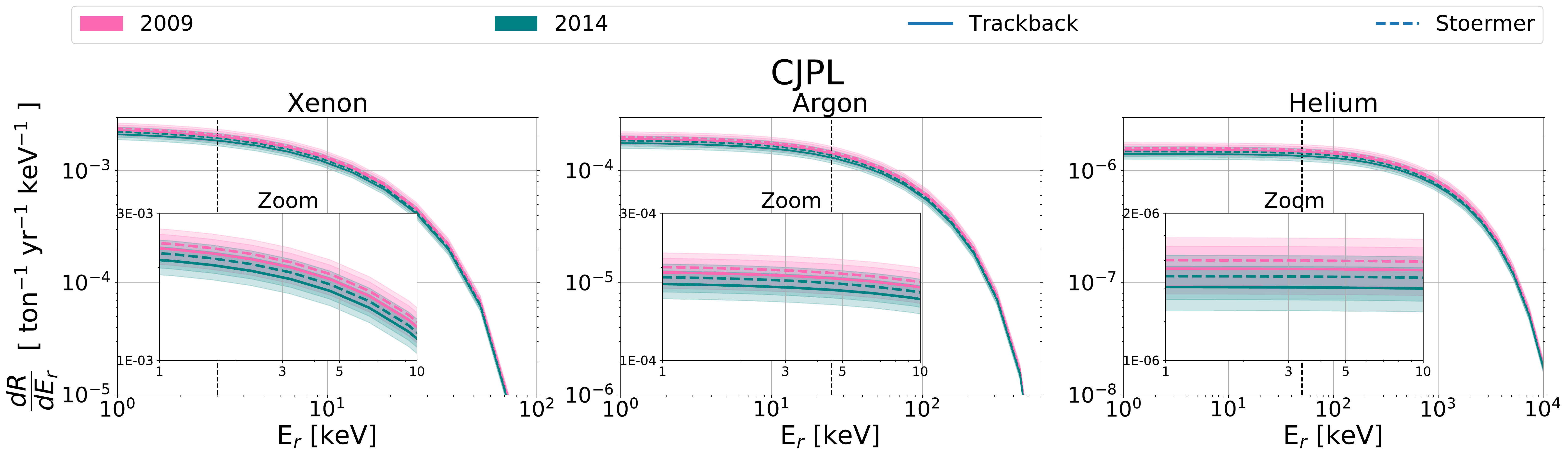}
    \includegraphics[width = 0.9\textwidth]{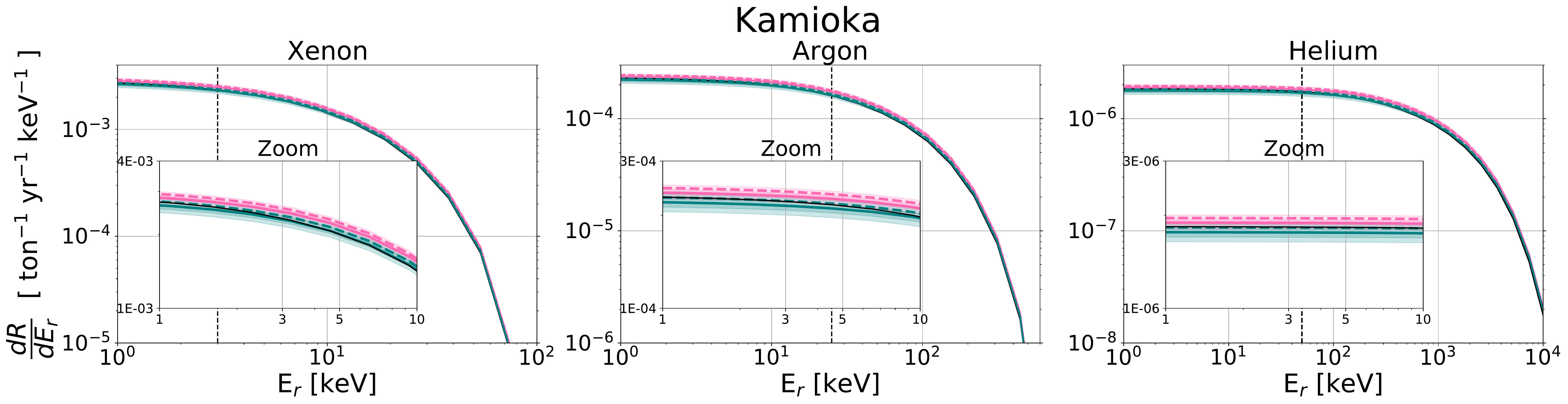}
    \includegraphics[width = 0.9\textwidth]{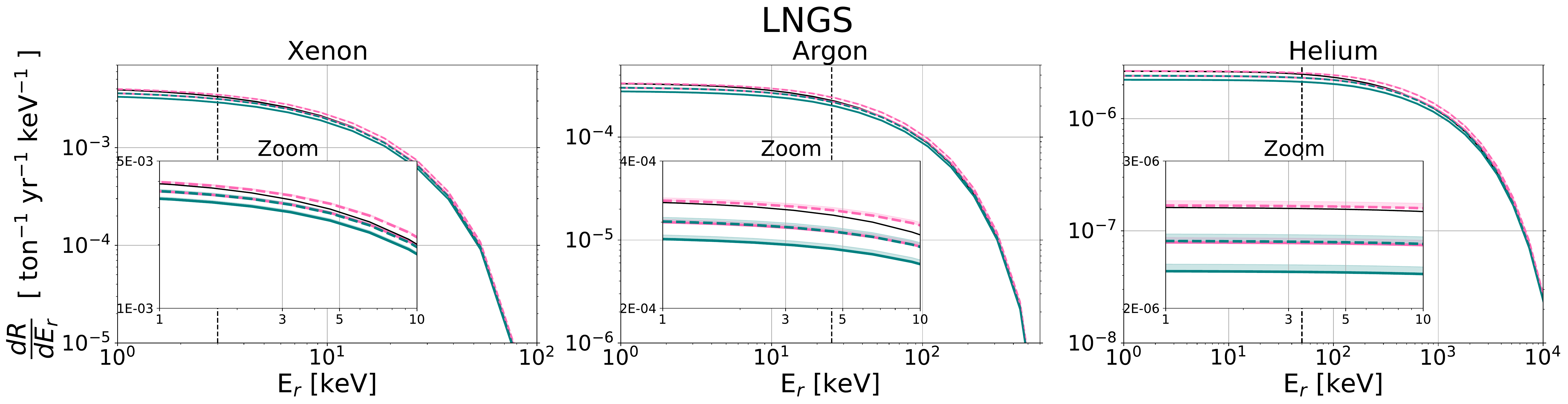}
    \includegraphics[width = 0.9\textwidth]{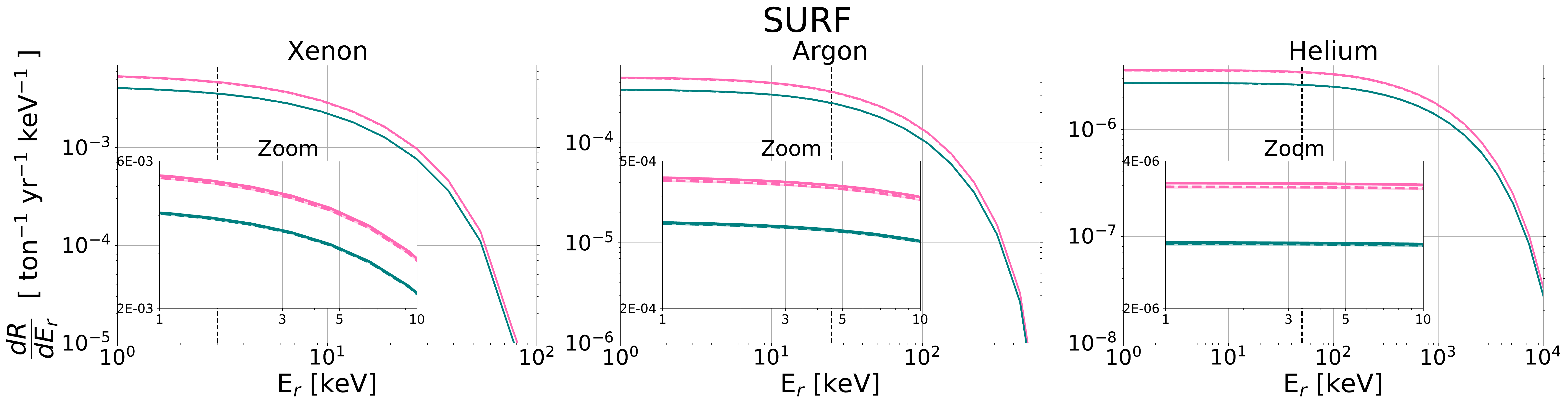}
    \includegraphics[width = 0.9\textwidth]{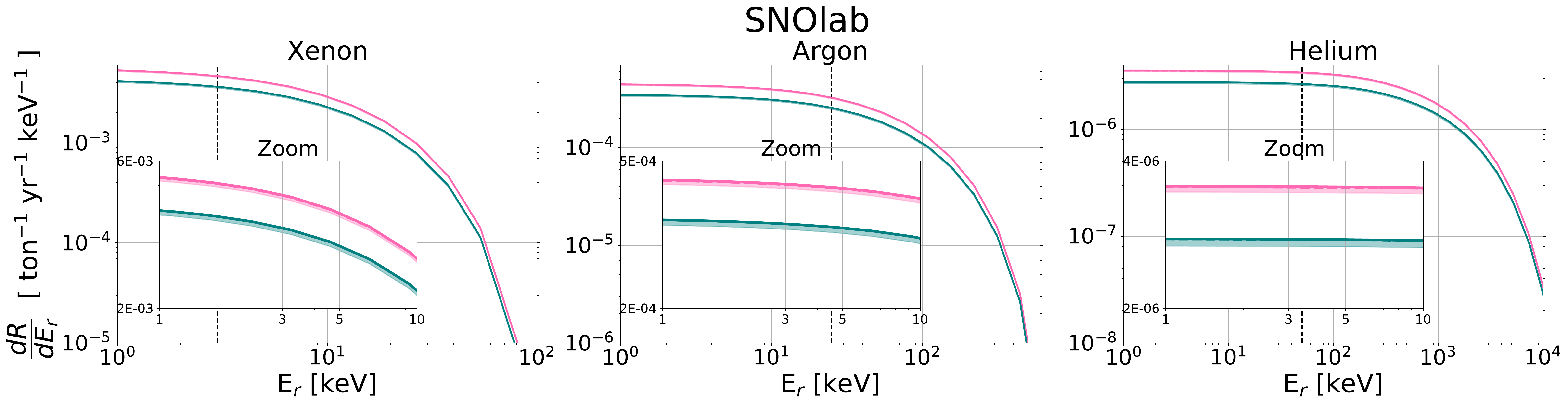}
    
    \caption{Differential event rate of nuclear recoil energies at CJPL, Kamioka, LNGS, SURF and SNO. Shown are the results for solar minimum (2009) and solar maximum (2014), and for the Stoermer and trackback geomagnetic models. The columns are for xenon, argon, and helium targets, respectively. The bands represent the uncertainty due to the assumed up-down ratio. Our benchmark energy thresholds are indicated as dashed vertical lines.}
    \label{fig:ton_keV_yr}
\end{figure*}

\begin{figure*}[!htbp]
    \centering
    \includegraphics[width = 0.9\textwidth]{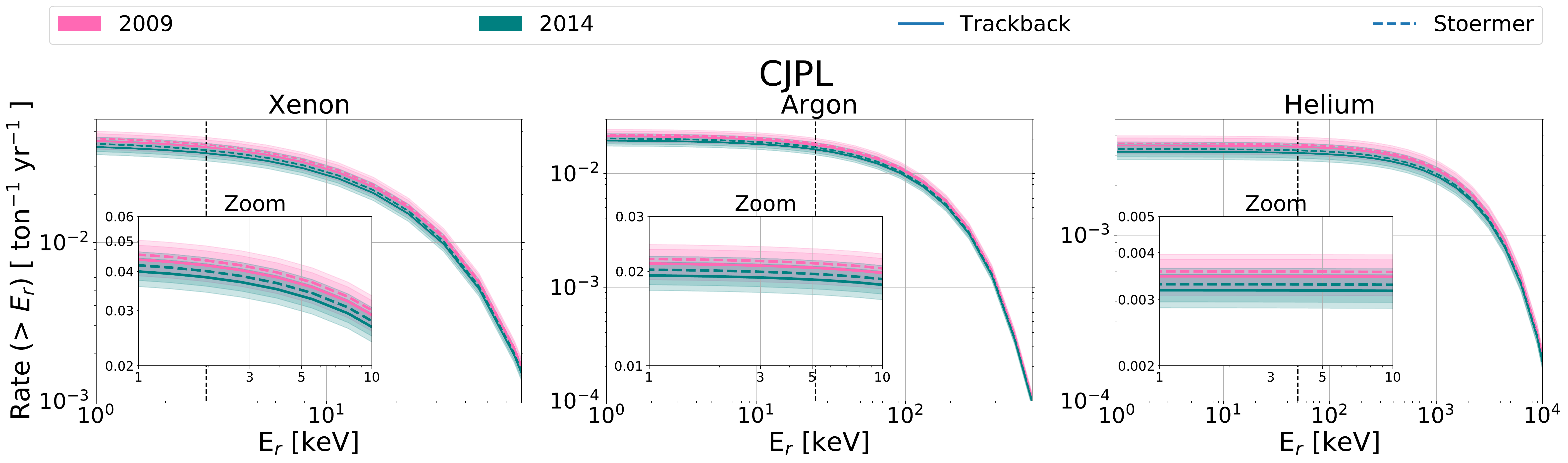}
    \includegraphics[width = 0.9\textwidth]{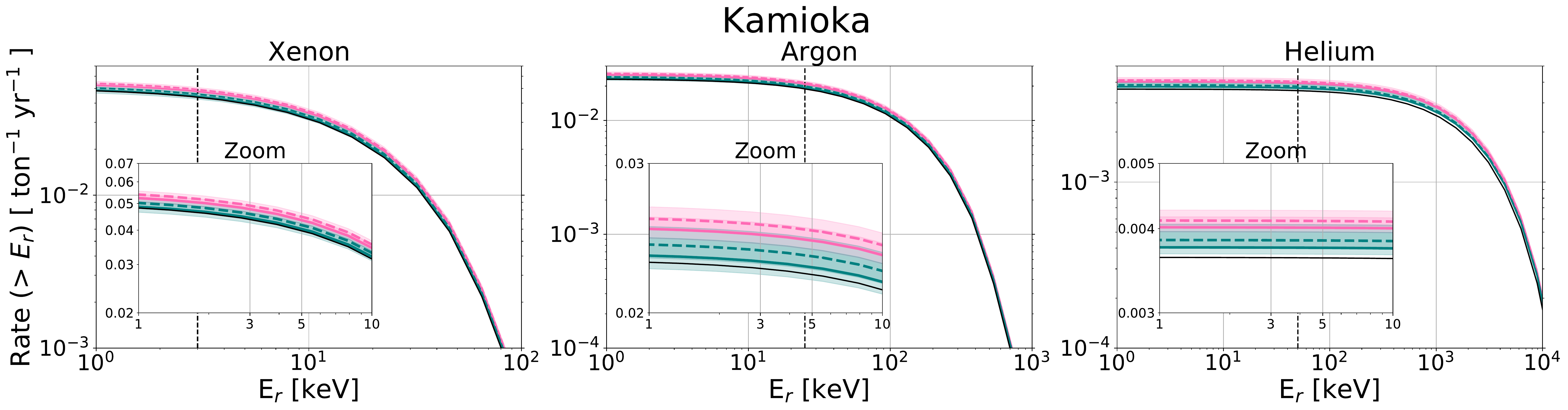}
    \includegraphics[width = 0.9\textwidth]{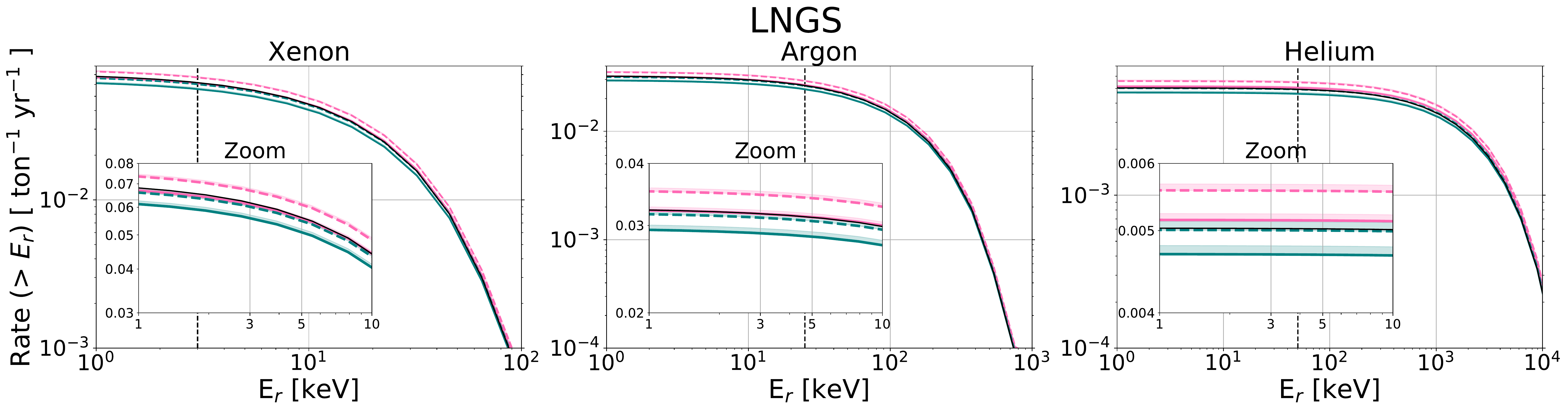}
    \includegraphics[width = 0.9\textwidth]{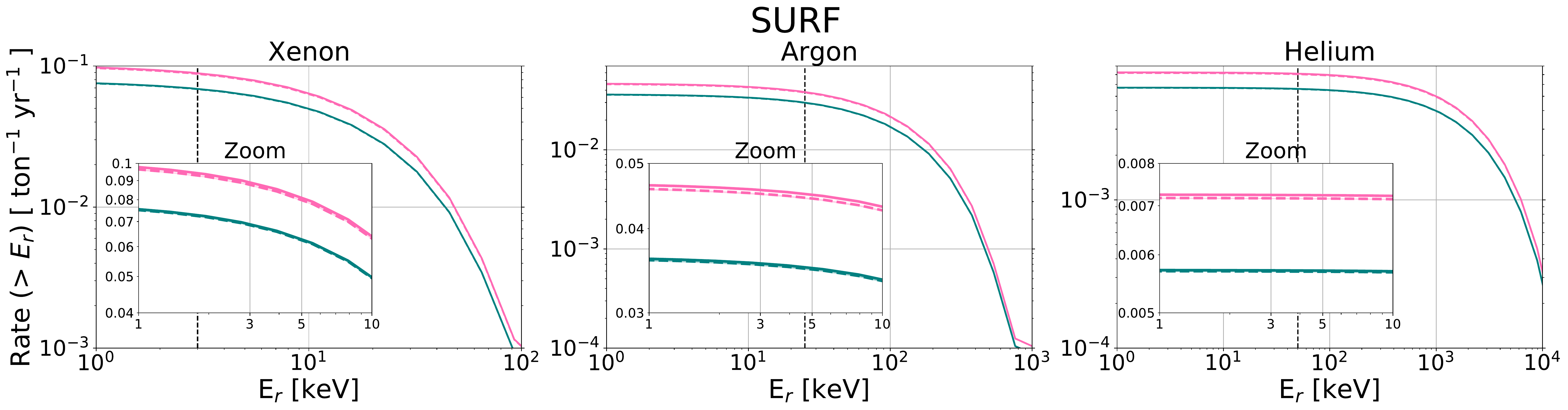}
    \includegraphics[width = 0.9\textwidth]{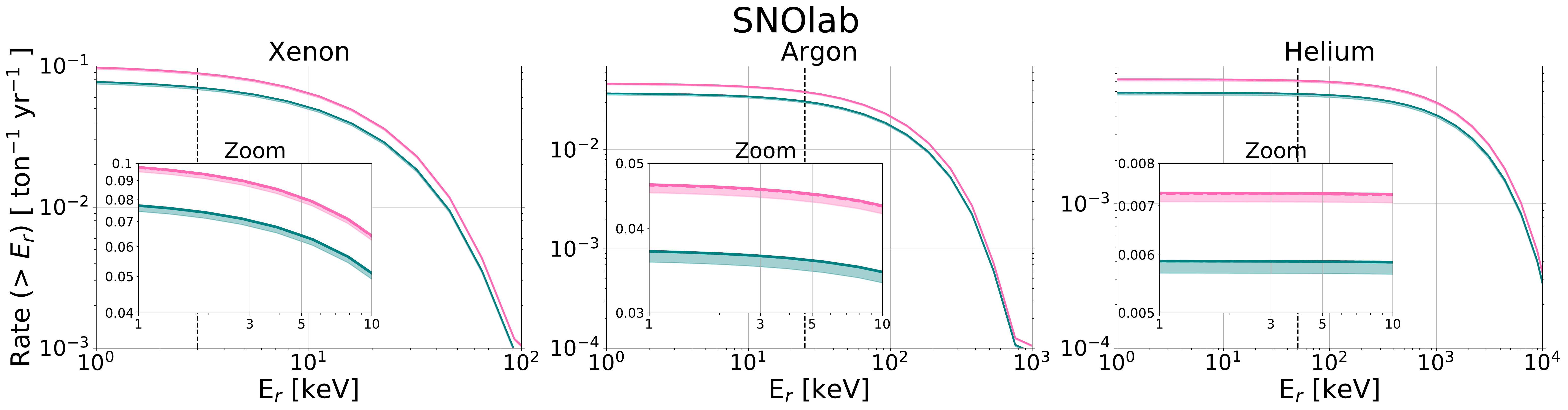}

    \caption{Integrated event rate of nuclear recoil energies above a given threshold energy $\mathrm{E_r}$ at CJPL, Kamioka, LNGS, SURF and SNO. Shown are the results for solar minimum (2009) and solar maximum (2014), and for the Stoermer and trackback geomagnetic models. The columns are for xenon, argon, and helium targets, respectively. The bands represent the uncertainty due to the assumed up-down ratio. Our benchmark energy thresholds are indicated as dashed vertical lines.}
    \label{fig:ton_yr}
\end{figure*}

\begin{figure*}[!htbp]
    \centering
    \includegraphics[width = \textwidth]{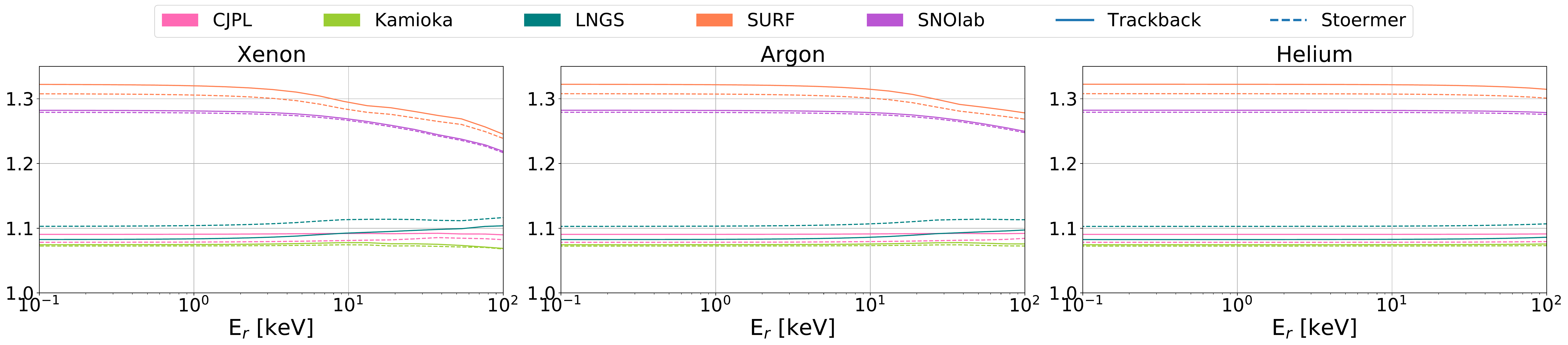}
    \caption{Ratio of the event rate at solar minimum to that at solar maximum, as a function of nuclear recoil energy. Shown are for different detector locations, and for the Stoermer and trackback geomagnetic models.}
    \label{fig:ratios}
\end{figure*}

\section{Results: Atmospheric neutrino event rates}
\label{sec:results} 
Combining the results for the flux with the scattering cross section, now we move on to predict the event rate for different detector targets. We first determine the range of neutrino energies that each nuclear mass target is sensitive to, at each detector location, and at each epoch in the solar cycle. Using Equation~\ref{eqn:eventrate_Er}, we show the distribution of neutrino energies for each of the three targets in Figure~\ref{fig: rate_vs_Enu}. For our canonical lower nuclear recoil energy thresholds, we assume \mbox{(3, 25, 50) keV} for (xenon, argon, helium), respectively. Shown are the calculations for both models described above for the rigidity cut-off. For all targets, we find that the shapes of these distributions are similar for both rigidity models, with the differences between the two models only discernible in the normalizations of the distributions. 

For each rigidity model, the bands reflect the uncertainty due to the up-down ratio that we assume, as defined for each detector location in Figure~\ref{fig:updownratio}. The uncertainty due to the up-down ratio is largest for detectors at low magnetic latitude, while it is the smallest for detectors at high magnetic latitude. The uncertainty due to the up-down ratio is particularly large at CJPL, because in this case we interpolate between the respective ratios at Kamioka and INO to obtain this ratio. 

Conversely, for detectors at high latitude, there is a discernible difference between the rate as measured at solar minimum relative to that at solar maximum. For example, for the SNO and SURF locations, there is an $\sim 30\%$ change in the flux from solar minimum to solar maximum. Further, for the high latitude locations like SNO and SURF, there is little to no dependence on the rigidity model assumed. The opposite is true for detectors at low latitude; at Kamioka and CJPL, there is $\lesssim 10\%$ change in the flux from solar minimum to solar maximum, but there is a larger change between the rigidity models assumed. 

As indicated in Figure~\ref{fig: rate_vs_Enu}, the lightest mass target, helium, is sensitive to the highest energy neutrinos. For comparison, the heaviest target, xenon, is sensitive to the lowest energy neutrinos in the distribution. This is due to our assumed detector energy thresholds, which are typical for current detectors using these target materials. However, it is interesting to note that for all of these targets, at all locations, these distributions peak at lower energy than the corresponding neutrino distributions that SK is sensitive to. This implies that a detection of these neutrinos would be the lowest energy atmospheric neutrinos yet detected. 

The distributions in Figure~\ref{fig: rate_vs_Enu} are sensitive to the assumed value of the nuclear recoil threshold energy. Though, as we discuss above, at the very lowest neutrino energies our atmospheric flux spectrum may under-predict the true flux, this is likely only an issue for the case of xenon with a low threshold of \mbox{3 keV}. The sensitivity of these distributions to nuclear recoil threshold energy in xenon is shown in Figure~\ref{fig: Xe_threshold}. 

To understand how reducing the nuclear energy threshold translates into the ability to access lower-energy neutrinos, in Figure~\ref{fig:trackbackxenon} we plot the normalized detector response function for xenon, for the case of solar minimum and solar maximum. At low neutrino energies, the response is relatively flat, in that a given neutrino energy maps onto a range of neutrino recoil energies. This indicates that there is some benefit to dropping the threshold to low energies in xenon, though neutrinos with energies $\sim 20$ MeV may still be accessed with a 1 keV threshold.

Figures~\ref{fig:ton_keV_yr} and~\ref{fig:ton_yr} show the differential and integrated event rates as a function of nuclear recoil threshold energy. Again in these figures the two rigidity models are shown as well as the predictions for solar minimum and solar maximum at each detector location. In each of these figures, we extend nuclear recoil energies up to the point at which the nuclear form factor begins to exponentially suppress the rate.

Figure~\ref{fig:ratios} is the ratio of event rate at solar minimum and solar maximum. In these figures, we see that the time variation of the flux is more significant at SNO and SURF, and is less discernible at CJPL and Kamioka. Given the low rate of atmospheric neutrino events in forseeable-future detectors, it will be challenging but possible to measure this time variation with sufficient statistics. From Figure~\ref{fig:ratios}, the flux ratios between solar minimum and solar maximum are maximally different at low recoil energies, and are reduced at higher recoil energies for all targets. This is primarily because the lower nuclear recoil energies preferentially sample the lowest energy portion of the atmospheric neutrino spectrum. 

\section{Discussion and Conclusions}
\label{sec:conclusions}

In this paper we have calculated the atmospheric neutrino event rate at a future dark matter detector for different phases in the solar cycle. Between solar minimum and solar maximum, we find that the flux normalization changes by the largest amount, $\sim 30\%$, for a detector at high magnetic latitude such as SNO or SURF. For a detector a lower magnitude latitudes, like CJPL, the flux variation between solar minimum and solar maximum is smaller, $\lesssim 10\%$. Further, detectors at high magnetic latitude are less sensitive to the assumed model for the geomagnetic field, which determines the rigidity cut-off of cosmic rays at a given detector location. On the other hand, low latitude locations are in turn less sensitive to the assumed value for the rigidity cut-off. 

The different amounts in time variation expected at different underground laboratories results from the region of the cosmic ray proton spectrum that is sampled at each location. At Kamioka or CJPL, regions of the proton spectrum $\sim 8-9$ GeV are sampled, and at these relatively high energies, time modulation is insignificant. At SNOlab and SURF proton energies down to $\sim 2-3$ GeV are sampled, where the time variation is most significant. At LNGS, intermediate proton energies down to $\sim 4-5$ GeV explain why the expected time modulation is between those at Kamioka/CJPL and SNOlab/SURF despite the geomagnetic field being similar to that at SNOlab/SURF. Sampling even lower cosmic ray energies would result in more significant expected time variation, but given the kinematics of the CE$\nu$NS channel, that would require lower detector energy thresholds than the ones considered here.

Though our simulation of the low-energy atmospheric neutrino flux has included several approximations as compared to previous simulations at higher energies, we have identified systematics associated with these approximations, and have attempted to quantify the uncertainties associated with them.  It is important to emphasize that the detection of atmospheric neutrinos at the energies we consider extend atmospheric neutrino studies to a new, low-energy regime that is yet unexplored with experiments and theory. Specifically, all of the nuclear targets that we consider (Xe, Ar, He) with feasible recoil thresholds sample neutrinos at lower energies than those that have been detected at Super-Kamiokande. 

As the theoretical calculations of the low-energy atmospheric neutrino flux are challenging, a possible experimental strategy to control systematics would involve placing one detector at low magnitude latitude and one detection at high magnetic latitude. Such detectors would sample the atmospheric neutrino distribution at different energies and different time variations. Given sufficient counting statistics, a comparison between two such locations could better constrain the flux normalization at this low energy regime of the atmospheric neutrino spectrum. 

As we have emphasized, detection of this time-varying signal would require both a detector to run for an extended time period, and an understanding of backgrounds that may affect the extraction of the atmospheric neutrino signal. A particularly key background is that from $pp$ solar neutrinos scattering on electrons~\citep{Newstead:2020fie}, which leak into the nuclear recoil band and thereby mimic an atmospheric neutrino interaction. Additional possible backgrounds come from diffuse supernova neutrinos, and from instrumental backgrounds associated with the detector. This makes a study of the detection sensitivity to this effect dependent on instrumental parameters of the particular detector under consideration. Because of these complications, we leave the detailed analysis of signal extraction in a future detector to future work. 

Because of capabilities of proposed detector technology, we have focused on target detectors that are sensitive only to energy depositions. If detectors that are sensitive to direction of the recoiling nucleus, there are unique and novel signatures of atmospheric neutrinos that may be studied at these energies. As an example, because of geometric effects, the flux peaks at large zenith angles, towards the horizon. Further, there is an east-west asymmetry in the predicted rate due to the rigidity cut-off, so that more particles travel from the magnetic west to the east that those that travel at the opposite direction. A (futuristic) detection of these neutrinos could provide new insight into the structure of the geomagnetic field and cosmic ray flux at low energy. 

While the focus of our study has been on the astrophysical systematics on the neutrino flux, if these systematics were controlled, this opened the possibility of using atmospheric neutrinos to search for new physics such as CP violation~\citep{Ghosh:2013yon} or non-standard neutrino interactions~\citep{Dutta:2020che}. Indeed, when including non-standard interactions, the systematics on the flux uncertainty could be even larger than what we discuss, so that atmospheric neutrinos may be an exciting means to identify new physics with multi-tonne scale dark matter experiments. 

\begin{acknowledgments}
YZ and LES acknowledge support from DOE Grant de-sc0010813, RLF from NSF grant PHY-2112803. The geomagnetic coordinates used in this paper (Table~\ref{tab:mag}) were provided by the WDC for Geomagnetism, Kyoto (http://wdc.kugi.kyoto-u.ac.jp/wdc/Sec3.html).
\end{acknowledgments}

\end{document}